\pdfoutput=1
\documentclass[12pt,a4paper]{article}

\usepackage{ifthen} 
\newboolean{pdflatex}
\setboolean{pdflatex}{true} 

\newboolean{articletitles}
\setboolean{articletitles}{true} 

\newboolean{uprightparticles}
\setboolean{uprightparticles}{false} 

\def\paperauthors{LHCb collaboration} 
\def\paperasciititle{Amplitude analysis of the D+_s -> pi- pi+ pi+ decay} 
\def\papertitle{Amplitude analysis of the $D^{+}_{s}\to\pi^{-} \pi^{+}\pi^{+}$ decay } 
\def\paperkeywords{{High Energy Physics}, {LHCb}} 
\def\papercopyright{\the\year\ CERN for the benefit of the LHCb collaboration} 
\def\paperlicence{CC BY 4.0 licence}
\def\paperlicenceurl{https://creativecommons.org/licenses/by/4.0/}

\usepackage{siunitx}
\usepackage{scalefnt}

\usepackage[top=1in, bottom=1.25in, left=1in, right=1in]{geometry}

\usepackage{microtype}
\usepackage{lineno}  
\usepackage{xspace} 
\usepackage{caption} 

\usepackage{graphicx}  
\usepackage{color}
\usepackage{colortbl}
\graphicspath{{./figs/}} 

\usepackage{amsmath} 
\usepackage{amssymb}
\usepackage{amsfonts}
\usepackage{upgreek} 

\newcommand*\patchAmsMathEnvironmentForLineno[1]{%
\expandafter\let\csname old#1\expandafter\endcsname\csname #1\endcsname
\expandafter\let\csname oldend#1\expandafter\endcsname\csname
end#1\endcsname
 \renewenvironment{#1}%
   {\linenomath\csname old#1\endcsname}%
   {\csname oldend#1\endcsname\endlinenomath}%
}
\newcommand*\patchBothAmsMathEnvironmentsForLineno[1]{%
  \patchAmsMathEnvironmentForLineno{#1}%
  \patchAmsMathEnvironmentForLineno{#1*}%
}
\AtBeginDocument{%
\patchBothAmsMathEnvironmentsForLineno{equation}%
\patchBothAmsMathEnvironmentsForLineno{align}%
\patchBothAmsMathEnvironmentsForLineno{flalign}%
\patchBothAmsMathEnvironmentsForLineno{alignat}%
\patchBothAmsMathEnvironmentsForLineno{gather}%
\patchBothAmsMathEnvironmentsForLineno{multline}%
\patchBothAmsMathEnvironmentsForLineno{eqnarray}%
}

\usepackage{hyperxmp}

\usepackage[pdftex,
            pdfauthor={\paperauthors},
            pdftitle={\paperasciititle},
            pdfkeywords={\paperkeywords},
            pdfcopyright={Copyright (C) \papercopyright},
            pdflicenseurl={\paperlicenceurl}]{hyperref}

\usepackage[colorinlistoftodos,textsize=scriptsize]{todonotes}

\usepackage[bottom,flushmargin,hang,multiple]{footmisc}

\usepackage[all]{hypcap} 

\usepackage{xspace} 
\usepackage{upgreek}

\def\lhcb   {\mbox{LHCb}\xspace}

\def\MagUp {\mbox{\em Mag\kern -0.05em Up}\xspace}

\ifthenelse{\boolean{uprightparticles}}%
{

 \def\Ppi         {\ensuremath{\uppi}\xspace}

 \def\Ppsi        {\ensuremath{\uppsi}\xspace}

 \def\PDelta      {\ensuremath{\Delta}\xspace}                 
 \def\PXi         {\ensuremath{\Xi}\xspace}                 
 \def\PLambda     {\ensuremath{\Lambda}\xspace}                 
 \def\PSigma      {\ensuremath{\Sigma}\xspace}                 
 \def\POmega      {\ensuremath{\Omega}\xspace}                 
 \def\PUpsilon    {\ensuremath{\Upsilon}\xspace}
 \let\oldPi\Pi
 \def\PPi         {\ensuremath{\oldPi}\xspace}

 \def\PB      {\ensuremath{\mathrm{B}}\xspace}                 
                  
 \def\PD      {\ensuremath{\mathrm{D}}\xspace}

 \def\PJ      {\ensuremath{\mathrm{J}}\xspace}                 
 \def\PK      {\ensuremath{\mathrm{K}}\xspace}

 \def\Pc      {\ensuremath{\mathrm{c}}\xspace}

 \def\Pi      {\ensuremath{\mathrm{i}}\xspace}

 \def\Ps      {\ensuremath{\mathrm{s}}\xspace}

 \def\thebaroffset{0.0em}
}
{

 \def\Ppi         {\ensuremath{\pi}\xspace}

 \def\Ppsi        {\ensuremath{\psi}\xspace}                 
                  
 \mathchardef\PDelta="7101
 \mathchardef\PXi="7104
 \mathchardef\PLambda="7103
 \mathchardef\PSigma="7106
 \mathchardef\POmega="710A
 \mathchardef\PUpsilon="7107
 \mathchardef\PPi="7105
                  
 \def\PB      {\ensuremath{B}\xspace}                 
                  
 \def\PD      {\ensuremath{D}\xspace}

 \def\PJ      {\ensuremath{J}\xspace}                 
 \def\PK      {\ensuremath{K}\xspace}

 \def\Pc      {\ensuremath{c}\xspace}

 \def\Pi      {\ensuremath{i}\xspace}

 \def\Ps      {\ensuremath{s}\xspace}

 \def\thebaroffset{0.18em}
}
\newcommand{\offsetoverline}[2][\thebaroffset]{\kern #1\overline{\kern -#1 #2}}%

\makeatletter
\ifcase \@ptsize \relax
  \newcommand{\miniscule}{\@setfontsize\miniscule{4}{5}}
\or
  \newcommand{\miniscule}{\@setfontsize\miniscule{5}{6}}
\or
  \newcommand{\miniscule}{\@setfontsize\miniscule{5}{6}}
\fi
\makeatother

\DeclareRobustCommand{\optbar}[1]{\shortstack{{\miniscule (\rule[.5ex]{1.25em}{.18mm})}
  \\ [-.7ex] $#1$}}

\def\squark    {{\ensuremath{\Ps}}\xspace}

\def\cquark    {{\ensuremath{\Pc}}\xspace}

\def\pion   {{\ensuremath{\Ppi}}\xspace}

\def\pip    {{\ensuremath{\pion^+}}\xspace}
\def\pim    {{\ensuremath{\pion^-}}\xspace}

\def\Kbar    {{\ensuremath{\offsetoverline{\PK}}}\xspace}
\def\Kb      {{\ensuremath{\Kbar}}\xspace}
\def\KorKbar {\kern \thebaroffset\optbar{\kern -\thebaroffset \PK}{}\xspace}

\def\D       {{\ensuremath{\PD}}\xspace}

\def\DorDbar {\kern \thebaroffset\optbar{\kern -\thebaroffset \PD}\xspace}

\def\Dp      {{\ensuremath{\D^+}}\xspace}
\def\Dm      {{\ensuremath{\D^-}}\xspace}

\def\DpDm    {\ensuremath{\Dp {\kern -0.16em \Dm}}\xspace}

\def\Ds      {{\ensuremath{\D^+_\squark}}\xspace}
\def\Dsp     {{\ensuremath{\D^+_\squark}}\xspace}

\def\B       {{\ensuremath{\PB}}\xspace}

\def\BorBbar {\kern \thebaroffset\optbar{\kern -\thebaroffset \PB}\xspace}

\def\Bd      {{\ensuremath{\B^0}}\xspace}

\def\BdorBdbar {\kern \thebaroffset\optbar{\kern -\thebaroffset \Bd}\xspace}

\def\Bs      {{\ensuremath{\B^0_\squark}}\xspace}

\def\BsorBsbar {\kern \thebaroffset\optbar{\kern -\thebaroffset \Bs}\xspace}

\def\jpsi     {{\ensuremath{{\PJ\mskip -3mu/\mskip -2mu\Ppsi}}}\xspace}

\def\Y#1S{\ensuremath{\PUpsilon{(#1S)}}\xspace}

\def\Lz          {{\ensuremath{\PLambda}}\xspace}

\def\LorLbar     {\kern \thebaroffset\optbar{\kern -\thebaroffset \PLambda}\xspace}

\def\Lc          {{\ensuremath{\Lz^+_\cquark}}\xspace}

\newcommand{\decay}[2]{\ensuremath{#1\!\to #2}\xspace} 

\def\to                 {\ensuremath{\rightarrow}\xspace}

\def\AT#1     {\ensuremath{A_{\mathrm{T}}^{#1}}\xspace}           

\def\C#1      {\ensuremath{\mathcal{C}_{#1}}\xspace}                       
\def\Cp#1     {\ensuremath{\mathcal{C}_{#1}^{'}}\xspace}                    
\def\Ceff#1   {\ensuremath{\mathcal{C}_{#1}^{\mathrm{(eff)}}}\xspace}        
\def\Cpeff#1  {\ensuremath{\mathcal{C}_{#1}^{'\mathrm{(eff)}}}\xspace}       
\def\Ope#1    {\ensuremath{\mathcal{O}_{#1}}\xspace}                       
\def\Opep#1   {\ensuremath{\mathcal{O}_{#1}^{'}}\xspace}                    

\newcommand{\aunit}[1]{\ensuremath{\text{\,#1}}}       

\newcommand{\tev}{\aunit{Te\kern -0.1em V}\xspace}
\newcommand{\gev}{\aunit{Ge\kern -0.1em V}\xspace}
\newcommand{\mev}{\aunit{Me\kern -0.1em V}\xspace}
\newcommand{\kev}{\aunit{ke\kern -0.1em V}\xspace}
\newcommand{\ev}{\aunit{e\kern -0.1em V}\xspace}
 
\newcommand{\mevc}{\ensuremath{\aunit{Me\kern -0.1em V\!/}c}\xspace}
\newcommand{\gevc}{\ensuremath{\aunit{Ge\kern -0.1em V\!/}c}\xspace}
\newcommand{\mevcc}{\ensuremath{\aunit{Me\kern -0.1em V\!/}c^2}\xspace}
\newcommand{\gevcc}{\ensuremath{\aunit{Ge\kern -0.1em V\!/}c^2}\xspace}

\def\fb   {\ensuremath{\aunit{fb}}\xspace}
\def\invfb   {\ensuremath{\fb^{-1}}\xspace}

\newcommand{\chisq}{\ensuremath{\chi^2}\xspace}

\newcommand{\chisqip}{\ensuremath{\chi^2_{\text{IP}}}\xspace}

\def\deriv {\ensuremath{\mathrm{d}}}

\def\gsim{{~\raise.15em\hbox{$>$}\kern-.85em
          \lower.35em\hbox{$\sim$}~}\xspace}
\def\lsim{{~\raise.15em\hbox{$<$}\kern-.85em
          \lower.35em\hbox{$\sim$}~}\xspace}

\def\sPlot{\mbox{\em sPlot}\xspace}

\def\pt         {\ensuremath{p_{\mathrm{T}}}\xspace}

\def\evtgen     {\mbox{\textsc{EvtGen}}\xspace}

\def\geant      {\mbox{\textsc{Geant4}}\xspace}

\def\photos     {\mbox{\textsc{Photos}}\xspace}

\def\pythia     {\mbox{\textsc{Pythia}}\xspace}

\def\tell1  {TELL1\xspace}
\def\ukl1   {UKL1\xspace}

\newcommand{\eg}{\mbox{\itshape e.g.}\xspace}

\newcommand{\lhcborcid}[1]{\href{https://orcid.org/#1}{\hspace*{0.1em}\raisebox{-0.45ex}{\includegraphics[width=1em]{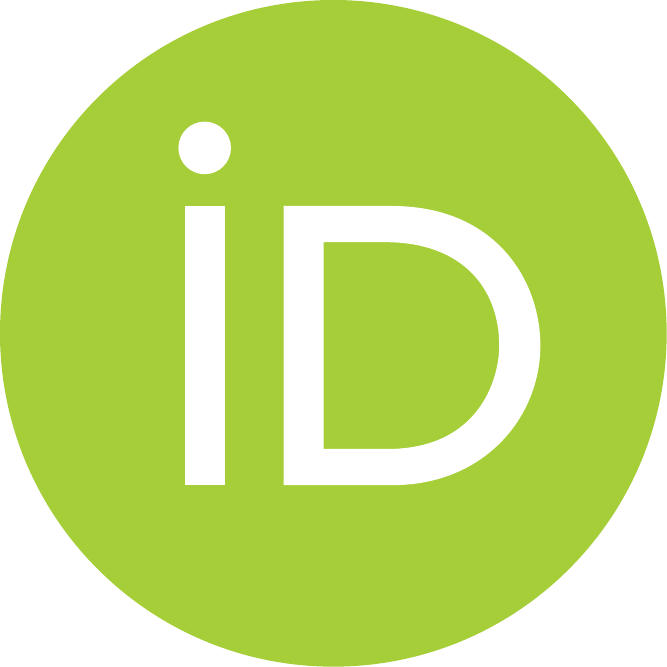}}}}

\def\Dppp         {\decay{\Dp}{\pim\pip\pip}}
\def\Dsppp        {\decay{\Ds}{\pim\pip\pip}}

\usepackage{cite} 
\usepackage{mciteplus}

\usepackage{longtable} 

\usepackage{booktabs} 
\usepackage{multirow}
\usepackage{siunitx}
\usepackage{subcaption}
\usepackage{placeins}
\usepackage{booktabs}
\usepackage{longtable}

\begin{document}

\renewcommand{\thefootnote}{\fnsymbol{footnote}}
\setcounter{footnote}{1}


\begin{titlepage}
\pagenumbering{roman}

\vspace*{-1.5cm}
\centerline{\large EUROPEAN ORGANIZATION FOR NUCLEAR RESEARCH (CERN)}
\vspace*{1.5cm}
\noindent
\begin{tabular*}{\linewidth}{lc@{\extracolsep{\fill}}r@{\extracolsep{0pt}}}
\ifthenelse{\boolean{pdflatex}}
{\vspace*{-1.5cm}\mbox{\!\!\!\includegraphics[width=.14\textwidth]{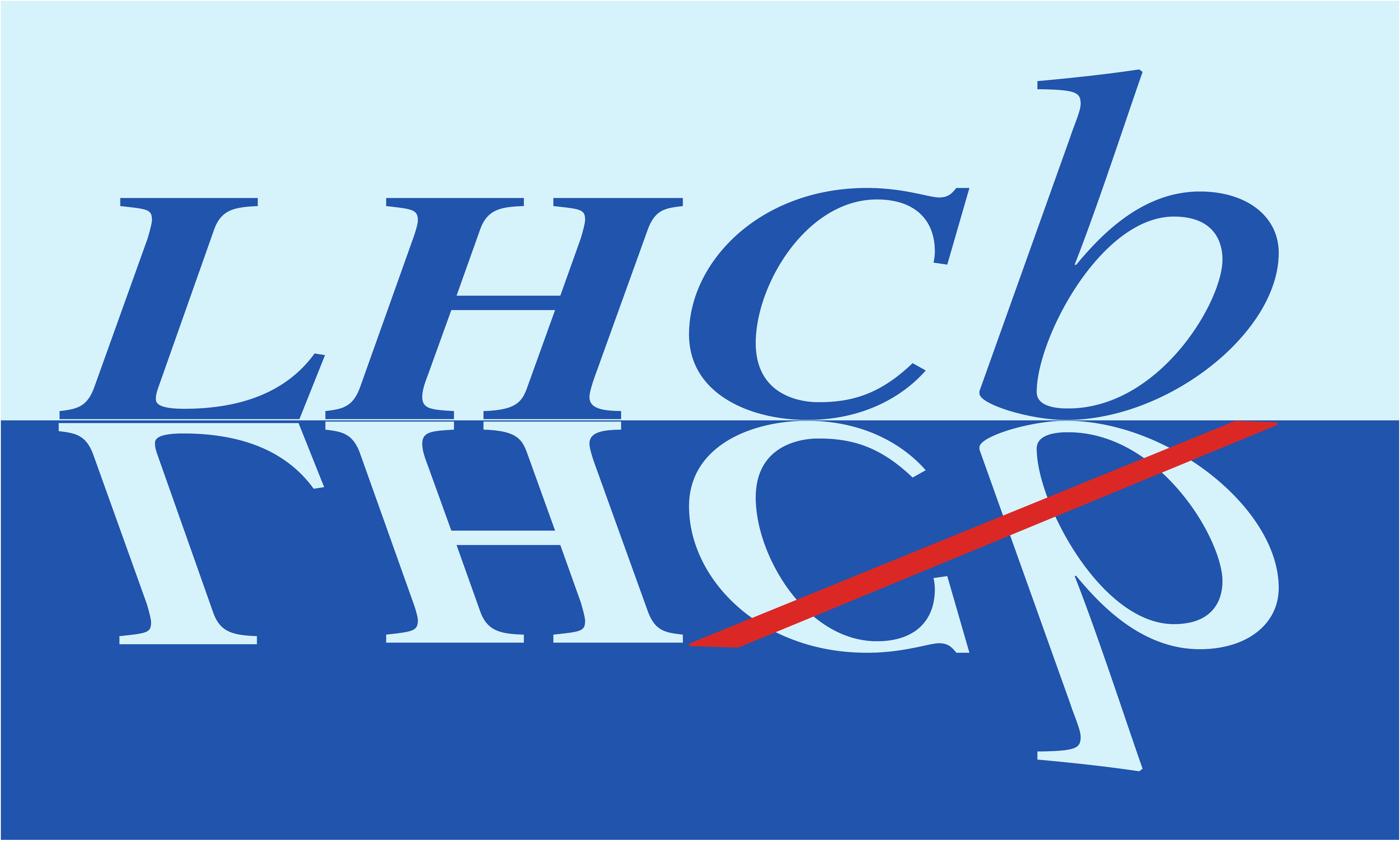}} & &}%
{\vspace*{-1.2cm}\mbox{\!\!\!\includegraphics[width=.12\textwidth]{figs/lhcb-logo.eps}} & &}%
\\
 & & CERN-EP-2022-173 \\  
 & & LHCb-PAPER-2022-030 \\  
 & & Oct 17, 2023 \\ 
 & & \\
\end{tabular*}

\vspace*{4.0cm}

{\normalfont\bfseries\boldmath\huge
\begin{center}
  \papertitle 
\end{center}
}

\vspace*{2.0cm}

\begin{center}
\paperauthors \footnote{Authors are listed at the end of this paper.}
\end{center}

\vspace{\fill}

\begin{abstract}
  \noindent
  A Dalitz plot analysis of the  \Dsppp decay is presented. The analysis is based on proton-proton collision data recorded by the LHCb experiment at a  centre-of-mass energy of 8\tev and corresponding to an integrated luminosity of 1.5\invfb. The resonant structure of the decay is obtained using a quasi-model-independent partial-wave analysis, in which the $\pi^+\pi^-$  S-wave amplitude is parameterised as a generic complex function determined by a fit to the data. The S-wave component is found to be dominant, followed by the contribution from spin-2 resonances and a small contribution from spin-1 resonances. The latter includes the first observation of the $\Dsp\to\omega(782)\pi^+$ channel in the \Dsppp decay. The resonant structures of the \Dsppp and \Dppp decays are compared, providing information about the mechanisms for the hadron formation in these decays.

\end{abstract}

\vspace*{2.0cm}

\begin{center}
   Published in JHEP 07 (2023) 204
\end{center}

\vspace{\fill}

{\footnotesize 
\centerline{\copyright~\papercopyright. \href{\paperlicenceurl}{\paperlicence}.}}
\vspace*{2mm}

\end{titlepage}


\newpage
\setcounter{page}{2}
\mbox{~}
%
%
%
%


\renewcommand{\thefootnote}{\arabic{footnote}}
\setcounter{footnote}{0}

\cleardoublepage


\pagestyle{plain} 
\setcounter{page}{1}
\pagenumbering{arabic}


\section{Introduction}
\label{sec:introduction}
Decays of $D$ mesons have unique features that can be explored for light-meson spectroscopy. In particular, their decays into three pseudoscalar particles proceed mainly through scalar, vector and tensor resonances, indicating that in these decays the dynamics of the final states is mainly driven by meson-meson interactions. This is the main motivation for the widely used isobar model in the analysis of \D-meson decays~\cite{asner}. The scalar mesons are the main component of decays into final states with two identical particles,  such as $D^+\to K^-K^+K^+$, $D^+\to K^-\pi^+\pi^+$, and  $D^+_{(s)}\to \pi^-\pi^+\pi^+$~\cite{PDG2022}. The amplitude analyses of these decays therefore offer an opportunity to access the S-wave $K\Kb$, $K\pi$ and $\pi\pi$ scattering amplitudes  from a single reaction with a well-defined initial state, starting from the corresponding invariant-mass threshold and continuing to $\sim$1.5--1.8 \gev. The information about meson-meson interactions provided by the analysis of $D$-meson decays is complementary to that from scattering experiments.

Although meson-meson scattering amplitudes play a key role in hadronic $D$-meson decays they cannot be measured directly~\cite{TripleM}. The $D$-meson decay is initiated by the short-distance weak transition of the charm quark. The hadrons are formed from the available quarks, and rescatter in all possible ways before reaching the detector. The final-state strong interactions between the decay products allow the scattering amplitudes to be assessed and, to a large extent, define the resonant structure of the final state. Unfortunately, at present it is not possible to describe all stages of the decay from first principles. Nevertheless, the measurement of the S-wave amplitude in different decay modes provides valuable inputs for phenomenological models.

In this paper, the resonant structure of the \Dsppp decay is determined.\footnote{Charge conjugation is implicit throughout this paper, unless otherwise stated. Natural units are used.} The analysis is based on proton-proton ($pp$) collision data collected by the LHCb experiment at a centre-of-mass energy of 8 \tev  corresponding to an integrated luminosity of 1.5 \invfb. The main purpose of this work is to determine the resonant structure of the decay and to measure the $\pi^+\pi^-$ S-wave amplitude. The results are obtained using a quasi-model-independent partial-wave analysis, in which the S-wave amplitude is parameterised as a generic complex function determined by a fit to the data. The resonant structure of the \Dsppp is compared to that of the \Dppp decay~\cite{LHCb-PAPER-2022-016}.

\begin{figure}[t] 
 \centering
\includegraphics[width= 0.5\textwidth]{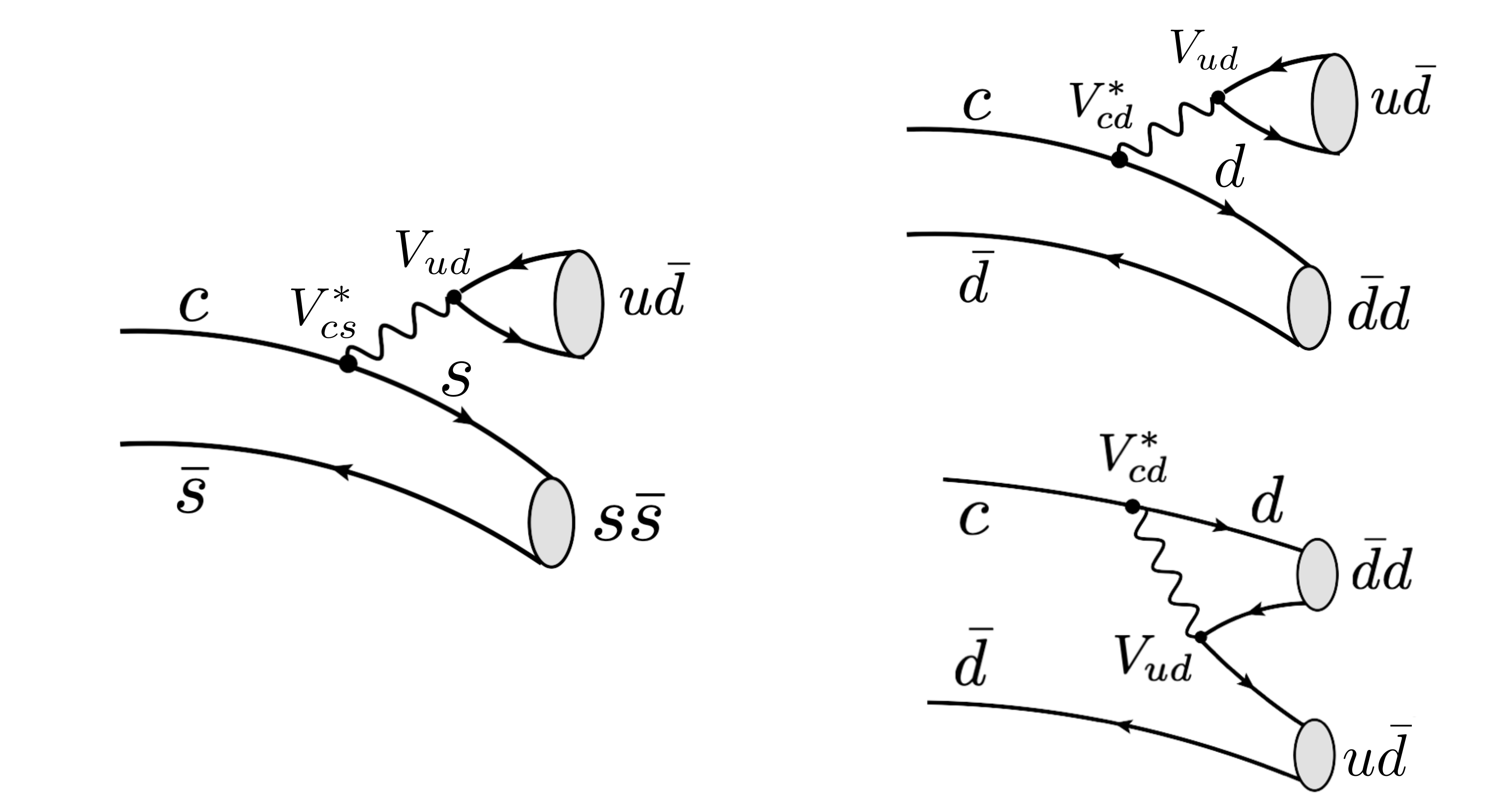}
\caption{Dominant tree-level diagrams leading to the (left) $D^+_s\to \pi^-\pi^+\pi^+$ and (right) \Dppp~decays. The resonances are produced from an $s\bar s$ pair in the  $D^+_s\to \pi^-\pi^+\pi^+$ decay and from a $d\bar d$ pair in the \Dppp~decay. For both decays, the annihilation diagram is suppressed.}
\label{fig:tree}
\end{figure}

A particular feature of the \Dsppp decay is that while  $D_s^+$ meson is a $c\bar s$ state and the decay occurs through the favoured transition $c\to s$, there are no strange particles in the final state. The \Dsppp decay is, therefore,  appropriate  for studies of resonances that couple to $\pi\pi$ and $K\Kb$ channels. This is the case of the $f_0(980)$, often interpreted as a non-$q\bar q$ state~\cite{PDG2022,Ochs_2013}. Previous analyses of the \Dsppp~\cite{babar-ds,besiiicollaboration2021amplitude} and \Dppp decays~\cite{e791Dp3pi,focus3pi,CleoD3pi} demonstrated that in both cases the dominant component is the S-wave amplitude, but with a different composition in each decay. In both cases, the main decay mechanism is expected to be the tree-level external $W$-emission amplitude, illustrated in Fig.~\ref{fig:tree}. In the \Dsppp decay, the  $c\to s$ transition results in an $s \bar s$ pair, from which the resonances arise. The  $D^+$ meson is a $c\bar d$ state and in the \Dppp decay the resonances are produced from a $d \bar d$ pair, resulting from the $c\to d$ transition. The different initial states at the quark level lead to different resonant structures despite the same final state. Therefore, the comparison between the resonant structures of the \Dppp and \Dsppp decays, and in particular between the $\pi^+\pi^-$ S-wave amplitudes, can improve our understanding of the mechanisms of hadron formation in nonleptonic decays of charm mesons.

%
\section{LHCb detector and simulation}
\label{sec:detector}

The LHCb detector~\cite{LHCb_detector2008,LHCb-DP-2014-002} is a single-arm forward spectrometer covering the pseudorapidity range $2 < \eta < 5$, designed for the study of particles containing \textit{$b$} or \textit{$c$} quarks. The detector includes a high-precision tracking system consisting of a silicon-strip vertex detector surrounding the $pp$ interaction region, a large-area silicon-strip detector located upstream of a dipole magnet with a bending power of about 4 Tm, and three stations of silicon-strip detectors and straw drift tubes placed downstream of the magnet. The tracking system provides a measurement of the momentum, $p$, of charged particles with a relative uncertainty that varies from 0.5$\%$ at low momentum to 1.0$\%$ at 200 \gev. The minimum distance of a track to a primary $pp$ collision vertex (PV), the impact parameter (IP), is measured with a resolution of ($15 + 29/\pt$) $\upmu$m, where \pt is the component of the momentum transverse to the beam, in Ge\kern -0.1em V. Different types of charged hadrons are distinguished using information from two ring-imaging Cherenkov detectors. Photons, electrons and hadrons are identified by a calorimeter system consisting of scintillating- pad and preshower detectors, an electromagnetic and a hadronic calorimeter. Muons are identified by a system composed of alternating layers of iron and multiwire proportional chambers. The online event selection is performed by a trigger that consists of a hardware stage, based on information from the calorimeter system, followed by a software stage, which applies a full event reconstruction. 

Simulations are used to model the effects of the geometrical acceptance of the detector and the selection requirements, and to evaluate efficiency variation across the Dalitz plot. In the simulation, $pp$ collisions are generated using \pythia~\cite{Sjostrand:2007gs,*Sjostrand:2006za} with a specific \lhcb configuration~\cite{LHCb-PROC-2010-056}. Decays of unstable particles are described by \evtgen~\cite{Lange:2001uf}, in which final-state radiation is generated using \photos~\cite{davidson2015photos}. The interaction of the generated particles with the detector, and its response, are implemented using the \geant toolkit~\cite{Allison:2006ve, Agostinelli:2002hh} as described in Ref.~\cite{LHCb-PROC-2011-006}.

\section{Candidate selection}
\label{sec:selection}

The \Dsppp decay candidates are selected online by a dedicated software trigger based on the decay topology. Requirements are also applied on the hardware trigger. In order to avoid distortions in the Dalitz plot of the \Dsppp decay, the analysis considers only events for which the hardware trigger decision was due to other particles not related to the signal. Three charged particles identified as pions according to particle-identification (PID) criteria are combined to form a good-quality decay vertex, detached from the associated PV, which is chosen as the PV with the smallest value of \chisqip. Here, \chisqip is defined as the difference in the vertex-fit \chisq of the PV reconstructed with and without the particle under consideration, in this case the  \Dsp candidate. Further requirements are applied on: the distance between the PV and the \Dsp decay vertex, the flight distance; the IP of the \Dsp candidate; the angle between the reconstructed \Dsp momentum vector and the vector from the PV to the decay vertex; the $\chi^2$ of the \Dsp decay vertex fit; the distance of closest approach between any two final-state tracks; and the momentum, the transverse momentum and the \chisqip of the \Dsp candidate and its decay products. The invariant mass of each \Dsp candidate is required to be within the interval \mbox{1910--2030\mev}.

Stringent PID requirements are applied to all three decay products, reducing to the per cent level the cross-feed from other charm-hadron decays such as $D^0\to K^-\pi^+$ with an unrelated track, $D^+_{(s)}\to \pi^-\pi^+\mu^+\nu_{\mu}$ and $\Lc\to p\pi^+\pi^-$. One important background from $D^+_s\to\eta'(\to \pi^+\pi^-\gamma)\pi^+$ decays, in which the photon is undetected, cannot be eliminated with PID requirements. This background affects mostly the region of the three-pion invariant mass region, $m(\pi^-\pi^+\pi^+)$, below the $D^+_s$ mass.

A multivariate analysis (MVA) is performed to further reduce the combinatorial background. The MVA uses the gradient boosted decision tree BDTG classifier~\cite{Breiman,AdaBoost}. Only the quantities related to the three-track combinations described above are used in the BDTG classifier, keeping the overall signal efficiency over the Dalitz plot as uniform as possible. The BDTG classifier is trained using simulated \Dsppp decays for the signal, and data from the $m(\pim\pip\pip)$ intervals  1920--1940\mev and 2010--2030\mev for the background. 

To improve the sample selection and the determination of the efficiency variation across the Dalitz plot, the simulation is weighted using the gradient boosted reweighter (GB-Reweighter) algorithm~\cite{rogozhnikov2016reweighting}. By applying the GB-Reweighter, residual differences between data and simulation are minimized. The weighting includes the kinematic distributions of the decay products. The target distributions are obtained from data with the \sPlot technique~\cite{Pivk:2004ty}.

\begin{figure}[h]
\centering
\includegraphics[width=0.6\textwidth]{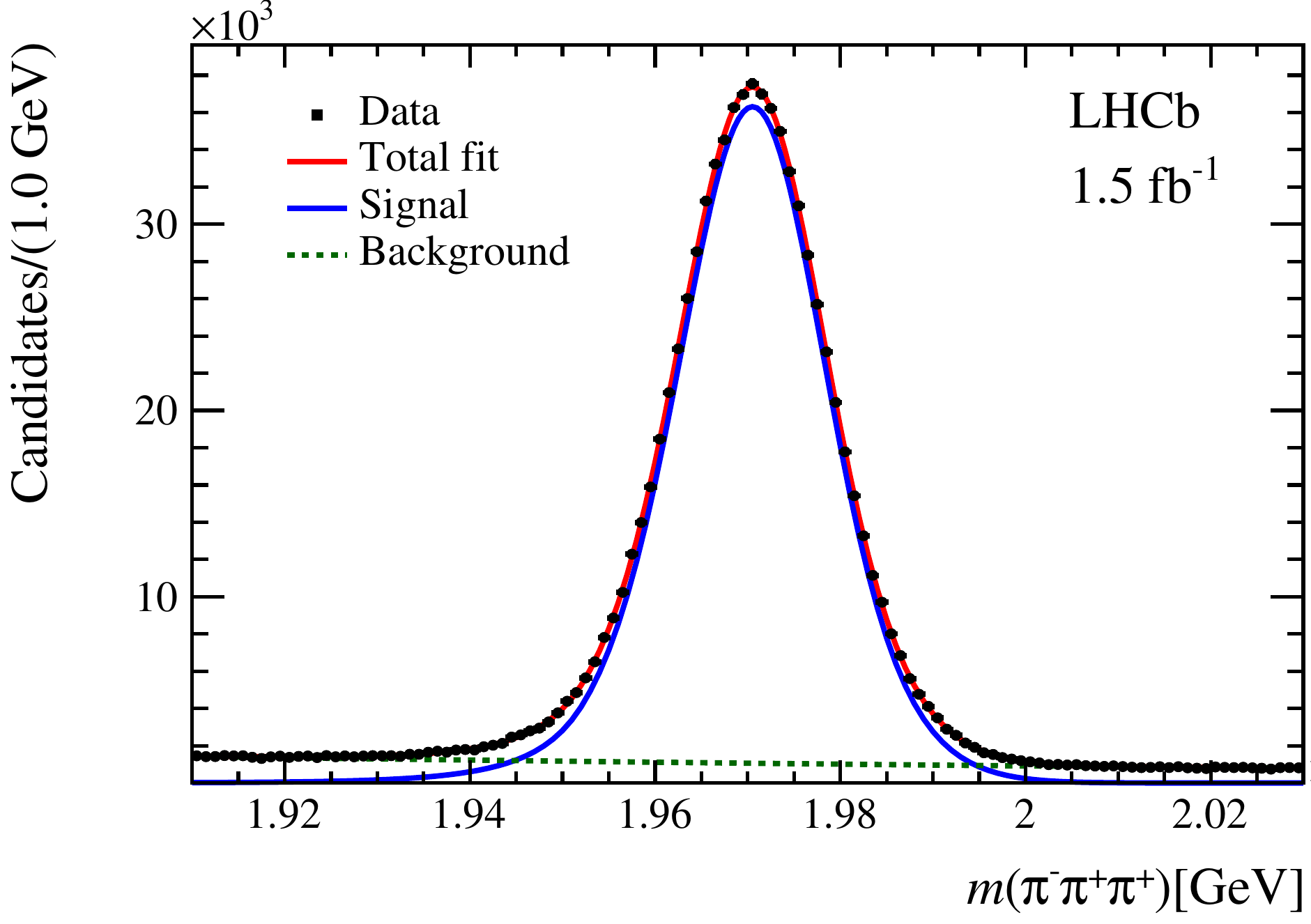}
\caption{Invariant-mass distribution of \Dsppp candidates after the final selection, with the fit result superimposed.}
\label{fig:final_3pimass}
\end{figure}

\begin{figure}[h]
\centering
\includegraphics[width=0.48\textwidth]{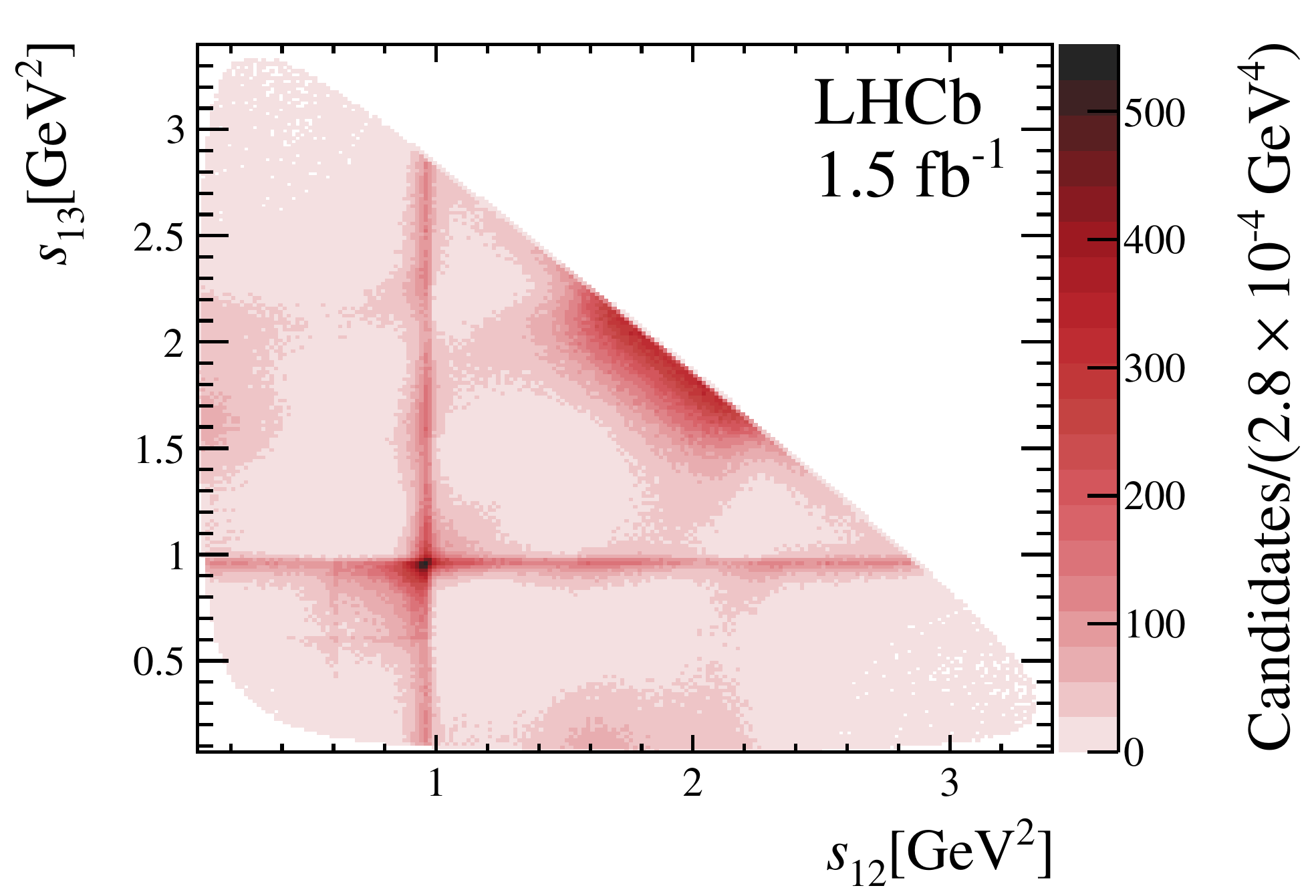}
\includegraphics[width=0.48\textwidth]{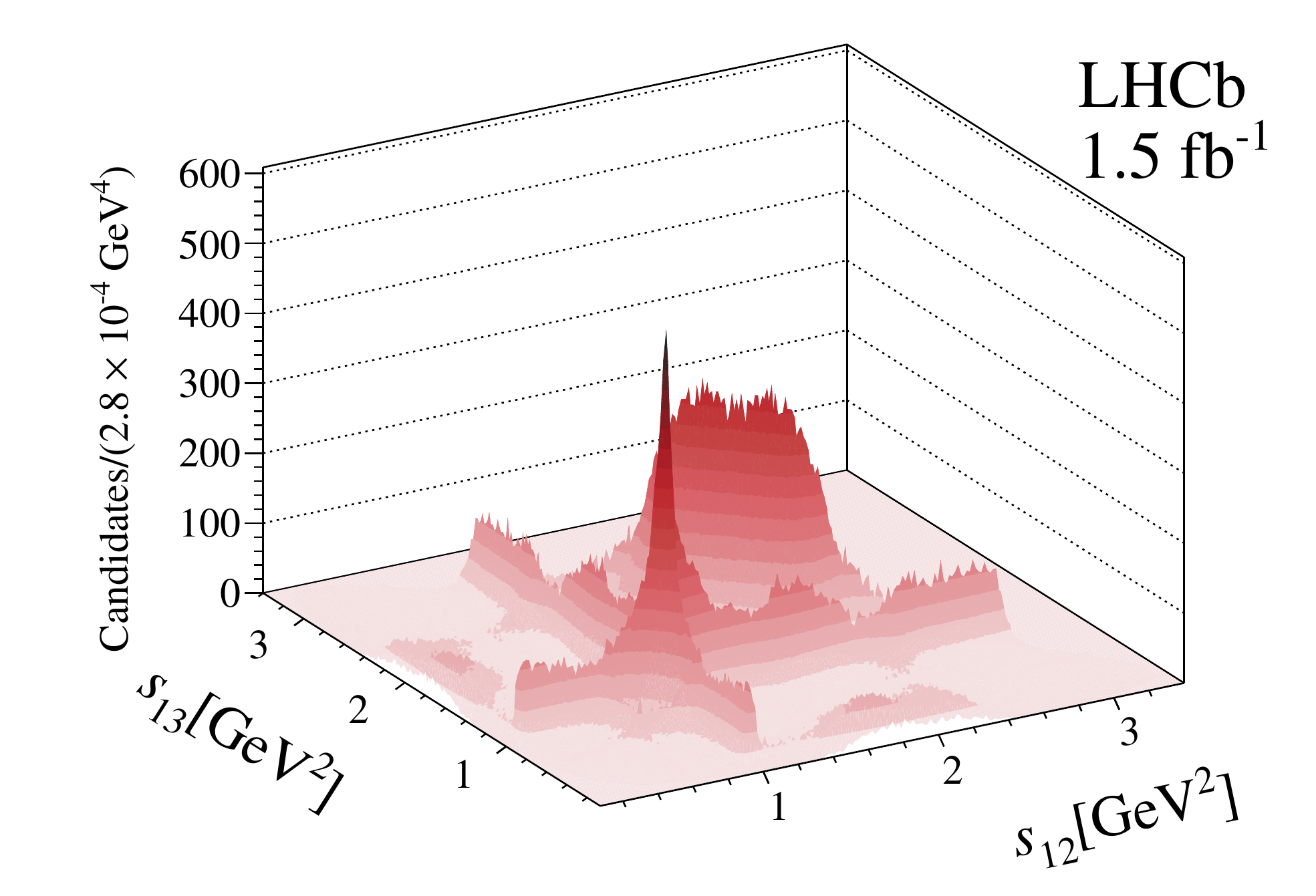}
\caption{(Top) Dalitz plot from selected \Dsppp candidates in the signal region. The colour scale indicates the density of candidates. (Bottom) Perspective view of the Dalitz plot.}
\label{fig:final_DP}
\end{figure}

The invariant-mass distribution of
\Dsppp candidates after the final selection is shown in Figure~\ref{fig:final_3pimass} with the fit result superimposed. The signal is represented by the sum of a Crystal Ball ~\cite{das2016simple} and 
a Gaussian function, with an effective width of  $\sigma_{\text{eff}}=8.9$\mev. The background is modeled by an exponential function. The Dalitz plot analysis is performed using candidates with $m(\pim\pip\pip)$ within $\pm2\sigma_{\text{eff}}$ of the known \Dsp mass~\cite{PDG2022}. The requirement on the BDTG output is chosen to yield a sample of \Dsppp decays with 95\% purity in this region. The efficiency drops rapidly for more stringent BDTG requirements, with only a modest gain in purity. The target purity is chosen to minimize the systematic uncertainty related to the background model. 
In approximately 0.5\% of the events there is more than one signal candidate, and all are retained for further analysis. The Dalitz plot of the selected candidates is shown in Fig.~\ref{fig:final_DP}.

\section{Signal efficiency and background model}
\label{sec:effandbkg}
The efficiency for reconstructing and selecting a \Dsppp candidate varies across the Dalitz plot.A combination of simulation and data-driven methods is used to determine this efficiency as a function of the Dalitz plot variables $s_{12}$ and $s_{13}$, defined in Section 5. Effects of the geometrical acceptance of the detector, reconstruction, trigger and selection are included in the simulation, except for the particle identification. The PID efficiency is determined from data calibration samples. Decays that can be reconstructed without particle identification, such as the $D^{*+}\to D^0(\to K^-\pi^+)\pi^+$ chain, are used to determine the efficiency as a function of momentum and pseudorapidity for each decay product. The PID efficiency of the candidate is the product of the efficiency for each final-state particle and is used to weight the simulated events that pass the remainder of the selection.

\begin{figure}[hbt]
\centering
    \begin{subfigure}[b]{0.48\textwidth}
        \centering
        \includegraphics[width=\textwidth]{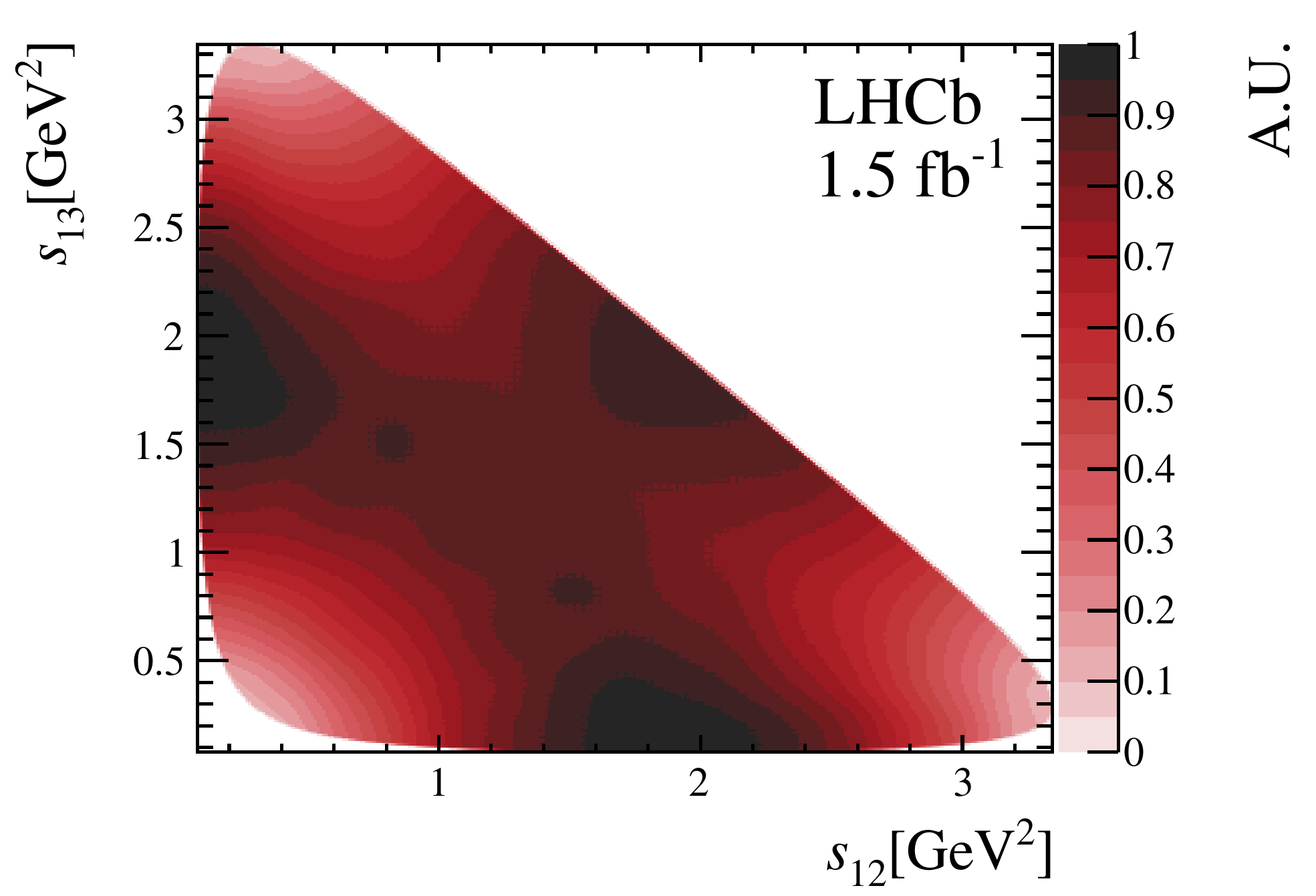}
        \label{fig:f1}
    \end{subfigure}
    \hfill
    \begin{subfigure}[b]{0.48\textwidth}
        \centering
        \includegraphics[width=\textwidth]{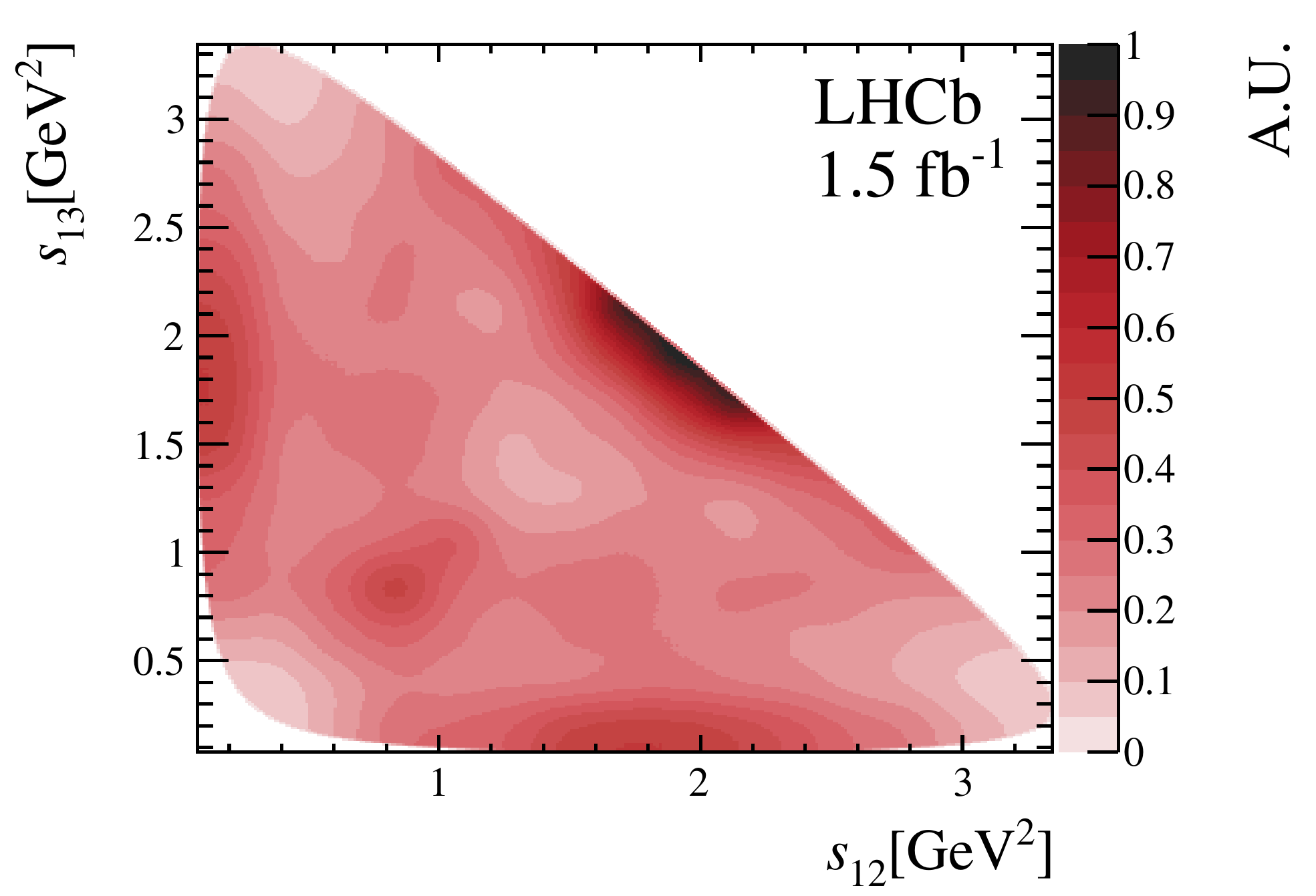}
        \label{fig:f2}
    \end{subfigure}
\caption{(Left) efficiency variation over the Dalitz plot  and (right) background  model  for \Dsppp decays.}
\label{fig:final_accep}
\end{figure}

The signal efficiency as a function of the Dalitz plot coordinates is a two-dimensional histogram with $15 \times 15$ uniform bins. This histogram is smoothed by a 2D cubic spline to avoid abrupt changes of the efficiency between neighboring bins. The smoothed efficiency histogram is shown in the left panel of Fig.~\ref{fig:final_accep}.

 The composition of the background varies across the $\pi^-\pi^+\pi^+$ invariant-mass spectrum. The background from other charm-hadron decays due to $K\to\pi$ misidentification is reduced to a negligible level by the stringent PID requirements, but these cannot eliminate contamination from the $D^+_s\to \eta'(\to \pi^+\pi^-\gamma)\pi^+$ decays. Since the photon is not detected, this background is present under the signal peak and extends to the lower sideband. In the upper sideband, the decay chain $D^{*+}\to D^0(\to \pi^-\pi^+)\pi^+$ is removed by vetoing candidates with a $\pi^+\pi^-$  mass greater than 1835 \mev.
 
 The background under the signal peak is modelled with a weighted average of data from the upper and lower sidebands corresponding to the intervals 1920--1940\mev and 2000--2020\mev, respectively.
 In the default result, equal weights are used for the two sidebands. Following the same procedure as for the efficiency, a histogram with $15 \times 15$ uniform bins is formed with candidates from the upper and lower sidebands, and smoothed by a 2D cubic spline. The final histogram is shown in the right panel of Fig.~\ref{fig:final_accep}.
\section{Formalism of the $\boldsymbol{\Dsppp}$ Dalitz plot fit}
\label{sec:formalism}

The \Dsppp decay amplitude  is defined as a coherent sum of the S-wave component ,$\mathcal{A}_{\text S}(s_{12}, s_{13})$, and individual contributions from the P- and D-wave resonant amplitudes,

 \begin{equation}
     \mathcal{A}\left(s_{12}, s_{13}\right)\equiv\mathcal{A}_{ S}\left(s_{12}, s_{13}\right)+\sum_i a_{i} e^{i \delta_{i}}\mathcal{A}_i\left(s_{12}, s_{13}\right)+\left(s_{12} \leftrightarrow s_{13}\right),
     \label{eq:fit}
 \end{equation}
\noindent where $s_{12}$ and $s_{13}$ are the invariant masses squared of the $\pi^-_1 \pi_2^+$ and $\pi^-_1 \pi_3^+$ pairs. The decay amplitude is symmetrised due to the presence of two identical particles in the final state. In the default fit, in addition to the S-wave component, six intermediate states are considered: the vector resonances $\rho(770)^0$, $\omega(782)$, $\rho(1450)^0$ and $\rho(1700)^0$; and the tensor resonances $f_2(1270)$ and $f'_2(1525)$. The $f_2(1270)\pi^+$ channel is chosen as reference, for which the magnitude and phase are set to 1 and 0, respectively. The magnitudes $a_i$ and relative phases $\delta_i$ of the the other resonant amplitudes are free parameters. The resonance parameters used in the fit are summarised in Table~\ref{tab:ressonances}.

The optimal values of the free parameters are obtained with an unbinned maximum-likelihood fit to the Dalitz plot. The likelihood function is defined as a combination of the signal and background probability density functions (PDFs), $\mathcal{P}_{\text{sig}}$ and $\mathcal{P}_{\text{bkg}}$,
\begin{eqnarray}
\mathcal{L}\equiv\prod_{j} \left[f_{\text {sig }}\times \mathcal{P}_{\text {sig }}(s_{12}^{j}, s_{13}^{j})+(1-f_{\text{sig}}) \times \mathcal{P}_{\text {bkg }}(s_{12}^{j}, s_{13}^{j})\right],
\end{eqnarray}
where $f_\mathrm{sig}$ is the fraction of the signal, fixed at the value obtained from the fit to the \pim\pip\pip mass spectrum, and $j$ runs over the candidates in the sample. The background PDF is determined from the sidebands, as described in Section~\ref{sec:effandbkg}. The set of optimal parameters is determined by minimising the quantity $-2\log \mathcal{L}$.

The normalised signal PDF is given by
\begin{align}
\mathcal{P}_{\mathrm{sig}}\left(s_{12}, s_{13}\right)&=\frac{1}{N_{\text {sig }}}\left|\mathcal{A}\left(s_{12}, s_{13}\right)\right|^{2} \epsilon(s_{12},s_{13}),\\
N_{\text {sig }}& = \int_{\rm DP}\left|\mathcal{A}\left(s_{12}, s_{13}\right)\right|^{2} \epsilon\left(s_{12}, s_{13}\right) \deriv s_{12} \deriv s_{13},
\label{sig-pdf}
\end{align}
where  $\epsilon(s_{12}, s_{13})$ is the signal efficiency and $N_{\text{sig}}$ is the normalisation integral over the Dalitz plot (DP).

The fit results are expressed in terms of the complex coefficients $a_i$ and  $\delta_i$ for each channel, and the corresponding fit fractions, $\mathrm{FF}_i$, which are computed by integrating the squared modulus of the corresponding amplitude over the phase space, and dividing by the integral of the total amplitude squared,
\begin{equation}
    \mathrm{FF}_{i}=\frac{\int_{\mathrm{DP}}\left|a_{i} e^{i\delta_i} \mathcal{A}_{i}\left(s_{12}, s_{13}\right)\right|^{2} \deriv s_{12} \deriv s_{13}}{\int_{\mathrm{DP}}\left|\mathcal{A}\left(s_{12}, s_{13}\right)\right|^{2} \deriv s_{12} \deriv s_{13}}~.
\end{equation}

The fit fractions and relative phases between the amplitudes are normalisation-independent quantities, allowing comparisons of results from different experiments. The sum of the fit fractions often differs from unity as a result of interference between the individual amplitudes. The interference fit fractions quantify the degree of interference between two particular components in the amplitude,
\begin{equation}
    \text{FF}_{i j}=\frac{\int_{\text{DP}} 2 \operatorname{Re}\left[a_{i}a_{j} e^{i(\delta_i-\delta_j)} \mathcal{A}_{i}\left(s_{12}, s_{13}\right) \mathcal{A}_{j}^{*}\left(s_{12}, s_{13}\right)\right] \deriv s_{12} \deriv s_{13}}{\int_{\mathrm{DP}}\left| \mathcal{A}\left(s_{12}, s_{13}\right)\right|^{2} \deriv s_{12} \deriv s_{13}}~.
\end{equation}
By construction, the sum of the fit fractions and interference terms is unity.
\begin{table}[h!]
    \centering
 
    \caption{Masses and widths of the resonances used in the fit~\cite{PDG2022}. Quoted uncertainties are used to estimate systematic uncertainties.} 
    \label{tab:ressonances}
\begin{tabular}{>{$}l<{$}| >{$}r<{\,\pm\,$}@{}>{$}l<{$} >{$}r<{\,\pm\,$}@{}>{$}l<{$}}
\toprule
\multicolumn{1}{c|}{Resonance} &
\multicolumn{2}{c}{$m_0~[\text{Me\kern -0.1em V}]$} &
\multicolumn{2}{c}{$\Gamma_0~[\text{Me\kern -0.1em V}]$} \\
\midrule
\rho(770)^{0}   & 775.26&0.23 & 149.1& 0.8  \\
\omega(782)     & 782.65&0.13 & 8.49 & 0.13 \\
\rho(1450)^{0}  & 1465  &25   & 400  & 60   \\
\rho(1700)^{0}  & 1720  &20   & 250  & 100  \\
f_2(1270)       & 1275.5&0.8  & 186.7& 2.2  \\
f'_2(1525)      & 1517.4&2.5  & 86   & 5    \\
\bottomrule
\end{tabular}
\end{table}

\subsection{Parameterisation of the S-wave amplitude}

The $\pi^+\pi^-$ S-wave amplitude is represented by a generic complex function determined from the data. The spectrum of the $\pim\pip$ invariant mass squared is divided into 50 intervals with approximately the same number of candidates. This approach ensures that narrower intervals are chosen where the amplitude varies rapidly, \eg around the $f_0(980)$ peak. At the lower edge of the interval $k$, the S-wave amplitude is determined by two real parameters, $\mathcal{A}_{S}^{k}\left(s_{\pi^{+} \pi^{-}}\right)=c_{k} e^{i \phi_{k}}$. Interpolations using one linear spline for the magnitude and one for the phase define the S-wave amplitude at any point of the spectrum. The set of 50 pairs $(c^k,\phi^k)$ are fit parameters. At any given point in the Dalitz plot with coordinates ($s_{12},s_{13}$) the amplitude is given by 
\begin{equation}
\mathcal{A}_{S}\left(s_{12}, s_{13}\right)=\mathcal{A}_{S}\left(s_{12}\right)+\mathcal{A}_{S}\left(s_{13}\right). 
\end{equation}

\subsection{Parameterisation of the P- and D-waves}

The amplitude for the resonant decay $D^+_s \rightarrow R \pi^+$,\ $R \rightarrow \pi^+ \pi^-$ is written as a product of form factors, $F_D$ and $F_R$, a function accounting for the angular distribution of the decay products, $\mathcal{M}_J(s_{12}, s_{13})$, and a dynamical function describing the resonance line shape, $T_{R}\left(s_{12}\right)$,
\begin{eqnarray}
\mathcal{A}_{i}\left(s_{12}, s_{13}\right)=F_D F_R\hskip .12cm \mathcal{M}_J(s_{12}, s_{13})\hskip .12cm T_{R}\left(s_{12}\right), \hskip .3cm s_{12}\leftrightarrow s_{13}~.
\end{eqnarray}

The terms $F_D$ and $F_R$ are the normalised Blatt--Weisskopf barrier factors~\cite{book:BlattWeisskopf} for the decays $D_s^+\to R \pi^+$ and $R\to \pi^-\pi^+$, respectively. The barrier factors account for the finite dimension of the particles involved in the reaction. They ensure the correct behaviour of the amplitude both at threshold and at the high end of the phase space, and depend on the orbital angular momentum $L$ of the decay products. Since the  $D_s^+$ meson is a spinless particle, the orbital angular momentum $L$ is equal to the resonance spin $J$. The normalised barrier factors are defined in terms of $z\equiv pr$, where $p$ is the modulus of the momentum of the decay products in the rest frame of the decaying particle, and $r$ is a parameter with dimension GeV$^{-1}$. The normalisation factor is defined in terms of $z_0 \equiv p_0r$, where $p_0$ is the decay momentum when the mass of the resonant system is equal to the known resonance mass. The values of the parameter $r$ are fixed at $r=r_{D} = 5.0\gev^{-1}$ for the transition $D_s^+ \rightarrow R \pi^+$, ~and $r=r_{R} = 1.5\gev^{-1}$ for the transition $R \rightarrow \pi^-\pi^+$. The formulae for the form factors are summarised in Table \ref{table:blatt}.

The dynamical functions $T_{R}(s_{ij})$ are parameterised by relativistic Breit--Wigner (RBW) functions, with the exception of the $\rho(770)^0$, for which the  Gounaris--Sakurai function~\cite{gounaris1968finite} is used. The relativistic Breit--Wigner function is 
\begin{eqnarray}
T_\mathrm{RBW}(s_{ij})=\frac{1}{m_{0}^{2}-s_{ij}-i m_{0} \Gamma(s_{ij})}~,
\end{eqnarray}

\begin{table}[h!] 
    \centering
    \caption{Spin-dependent Blatt--Weisskopf barrier factors. The normalisation ensures that the barrier factors are equal to unity at the resonance mass.\label{table:blatt}}
    \centering
\begin{tabular}{c|c}
\toprule
Resonance spin & Barrier factor\\
\midrule
1 & $(1 + z_0^2)^{1/2}\times(1 + z^2)^{-1/2}$  \\ 
2 & $(z_0^4 + 3z_0^2 + 9)^{1/2}\times(z^4 + 3z^2 + 9)^{-1/2}$  \\
\bottomrule
\end{tabular}
\end{table} 

\noindent
where $m_0$ and $\Gamma(s_{ij})$ are the known resonance mass and mass-dependent width, respectively. The mass-dependent width is expressed as
\begin{eqnarray}
\Gamma\left(s_{ij}\right)=\Gamma_{0}\left(\frac{p}{p_{0}}\right)^{2 J+1} \frac{m_{0}}{\sqrt{s_{ij}}} F_{R}^{2}(z),
\end{eqnarray}
where $\Gamma_0$ is the known value of the resonance width. The values $m_0$ and $\Gamma_0$ for all resonances are fixed in the fit.

The $\omega(782)$ is a narrow resonance and the finite mass resolution of the detector cannot be neglected. The mass resolution in the region around the $\omega(782)$ mass is 2.3\mev and it is accounted for by a convolution of the $\omega(782)$ Breit--Wigner distribution with a Gaussian function~\cite{ABRAROV20111894}.

The Gounaris--Sakurai function, a modification of the RBW commonly used to describe the pion electromagnetic form factor, is given by
\begin{equation}
T_\mathrm{GS}(s_{ij}) =  \frac{1+\Gamma_0d/m_0}{(m_0^2 - s_{ij}) + f(s_{ij}) - i m_0 \Gamma(s_{ij})}, 
\end{equation}
where
 \begin{equation}
\begin{split}
f(s_{ij}) &= \Gamma_0 \,\frac{m_0^2}{p_0^3}\,\left[\; p^2 \left(h(s_{ij})-h_0\right) +
       \left(\,m_0^2-s_{ij}\,\right)\,p^2_0\,
       \frac{\deriv h}{\deriv s_{ij}}\bigg|_{m_0}
       \;\right]~. \\
    \end{split}
\end{equation}

The parameter $h_0$ is the value of the function $h(s_{ij})$ when $s_{ij}=m_0^2$. The function $h(s_{ij})$ is given by
\begin{equation}
    h(s_{ij}) = \frac{2}{\pi}\,\frac{p}{\sqrt{s_{ij}}}\,
       \ln\left(\frac{\sqrt{s_{ij}}+2p}{2m_\pi}\right)~,
\end{equation}
where $m_{\pi}$ is the pion mass with 
\begin{equation}
\frac{\deriv h}{\deriv s_{ij}}\bigg|_{m_0} =
h_0\left[(8p_0^2)^{-1}-(2m_0^2)^{-1}\right] \,+\, (2\pi m_0^2)^{-1}~.
\end{equation}

The parameter $d=f(0)/(\Gamma_0 m_0)$ is given by
\begin{equation}
  d = \frac{3}{\pi}\frac{m_\pi^2}{p_0^2}
  \ln\left(\frac{m_0+2p_0}{2m_\pi}\right)
  + \frac{m_0}{2\pi\,p_0}
  - \frac{m_\pi^2 m_0}{\pi\,p_0^3}~.
\end{equation}

The Lorentz-invariant functions $\mathcal{M}_J$ describe the angular distribution of the decay products, accounting for the conservation of angular momentum. They are obtained using the covariant tensor formalism~\cite{asner}. For the general decay $D\to R c$, $R\to ab$, the explicit form of the functions $\mathcal{M}_J$ are
\begin{equation}
\mathcal{M}_{1} = s_{bc} - s_{ab} + \left(\frac{1}{s_{ab}} (m_D^2 - m_c^2) (m_a^2 - m_b^2)\right)~,
\label{spin1}
\end{equation}
for spin-1 resonances, and

\begin{align}
\mathcal{M}_{2}&= \mathcal{M}_{1}^2 - \frac{1}{3}\left(s_{ab} - 2m_D^2 -m_c^2 + \frac{1}{s_{ab}}(m_D^2-m_b^2)^2\right) \nonumber \\ &\times \left(s_{ab} - 2m_a^2 - 2m_c^2 + \frac{1}{s_{ab}}(m_a^2-m_b^2)^2 \right)~,
\label{spin2}
\end{align}
for spin-2 resonances.

\subsection{Goodness-of-fit}
To assess the goodness-of-fit, the Dalitz plot is divided into $n_b=1600$ bins containing approximately 500 candidates, and a $\chi^2$ test statistic is computed. The test statistic is defined as
\begin{equation}
    \chi^{2}\equiv\sum_{i=1}^{n_{b}} \chi^{2}_i =\sum_{i=1}^{n_{b}} \frac{(N_i^{\rm fit}-N_i^{\rm obs})^2}{N_i^{\rm obs}}~,
\end{equation}
where $N_i^\mathrm{obs}$ and  $N_i^\mathrm{fit}$ are the observed number of candidates and the population  estimated by the fit in bin $i$, respectively. The quantity $\chi^2/\textrm{ndof}$ is used as an estimator of the fit quality. The number of degrees of freedom, ndof, is in the range $[n_b-n_p-1, n_b-1]$, with $n_p = 110$ being the number of free parameters. The distribution of the normalised residuals $(N_i^{\rm fit}-N_i^{\rm obs})/\sqrt{N_i^{\rm obs}}$ is also used as an indication of the fit quality.  

\section{Results}
\label{sec:results}

\begin{figure}[h!]
    \centering
    \includegraphics[width = 0.45\textwidth]{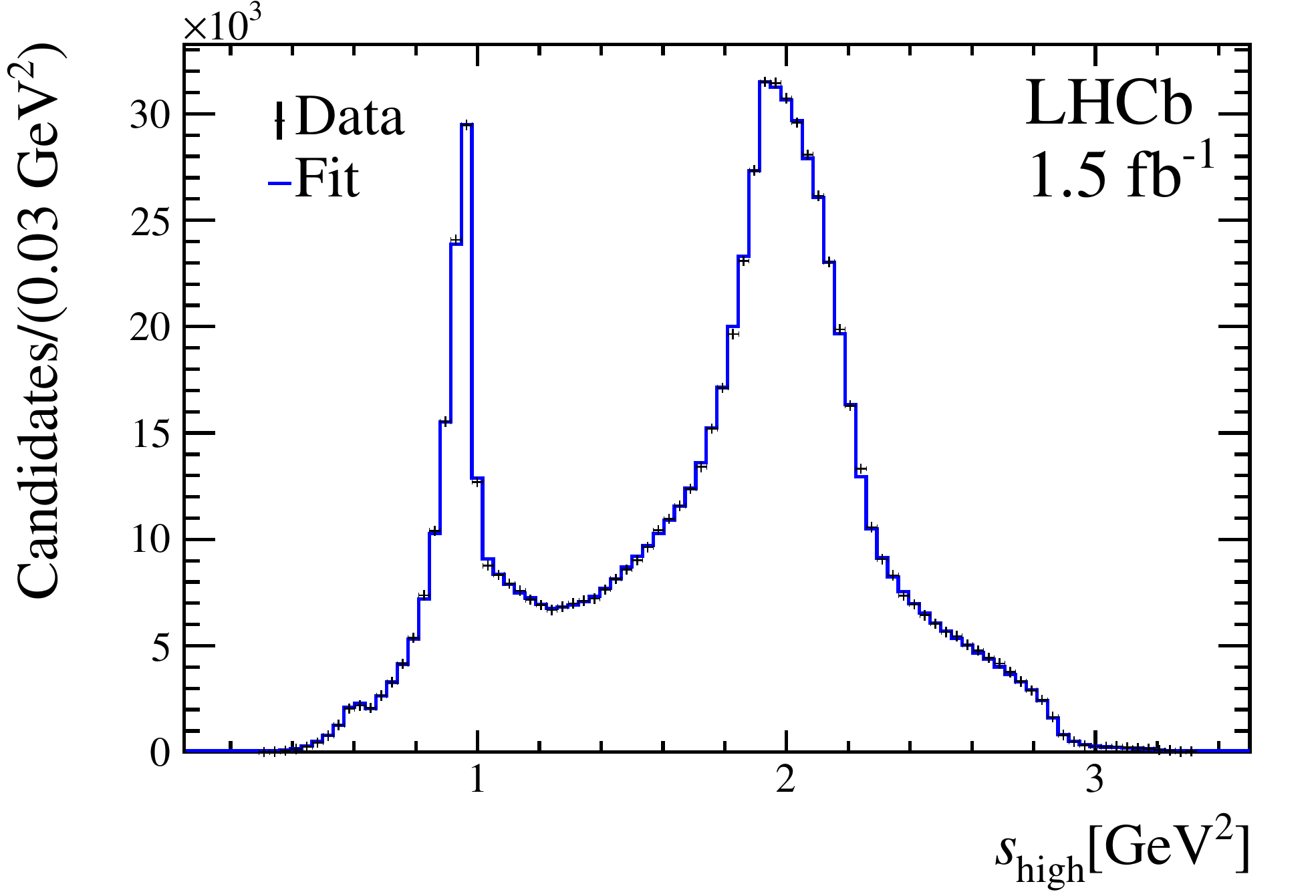}\includegraphics[width = 0.45\textwidth]{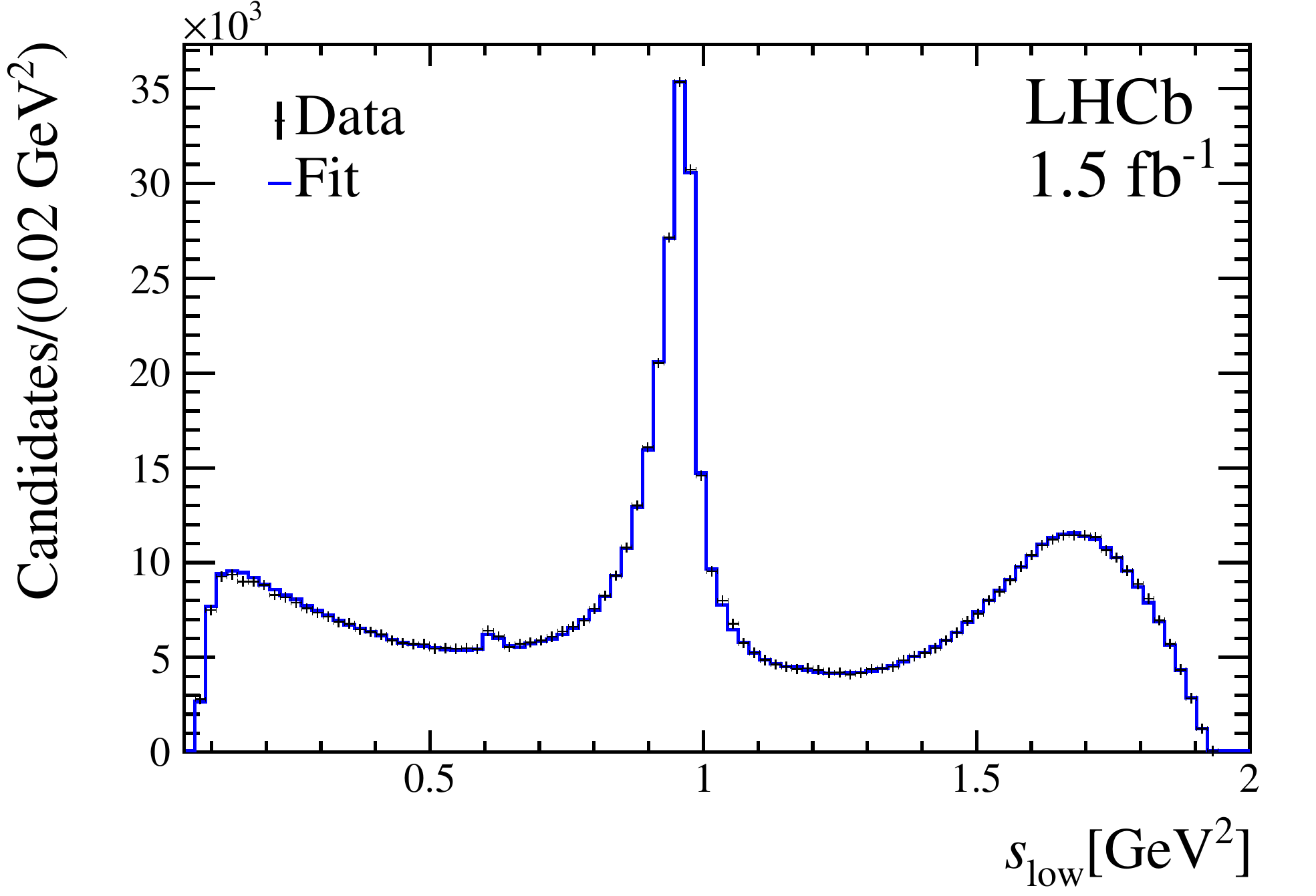}
    \caption{ {The folded Dalitz plot projections (left) $s_{high}$ and (right) $s_{low}$}. }
    \label{fig:fitPWA1a}
\end{figure}
\begin{figure}[h!]
    \centering
    \includegraphics[width = 0.45\textwidth]{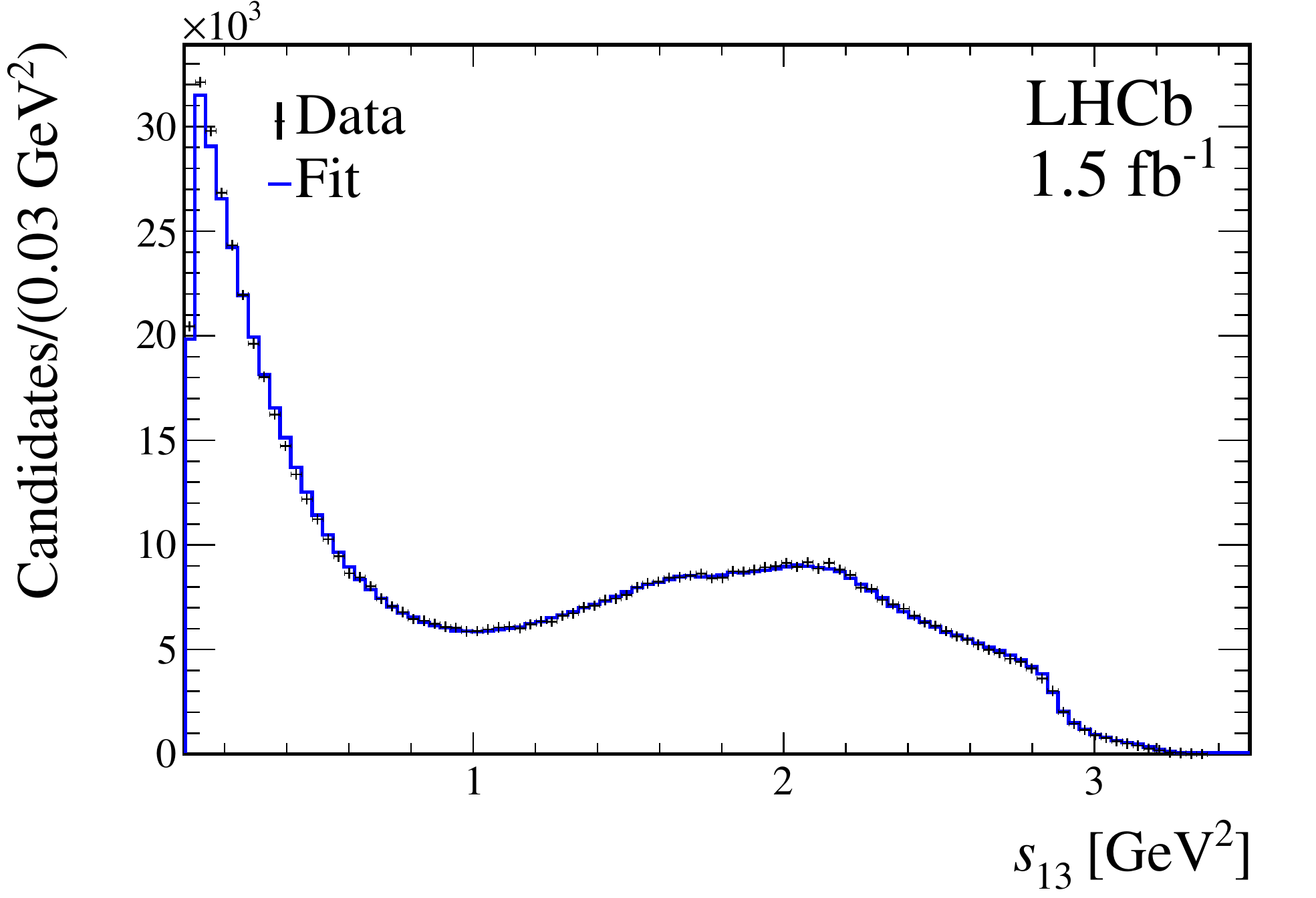}\includegraphics[width = 0.45\textwidth]{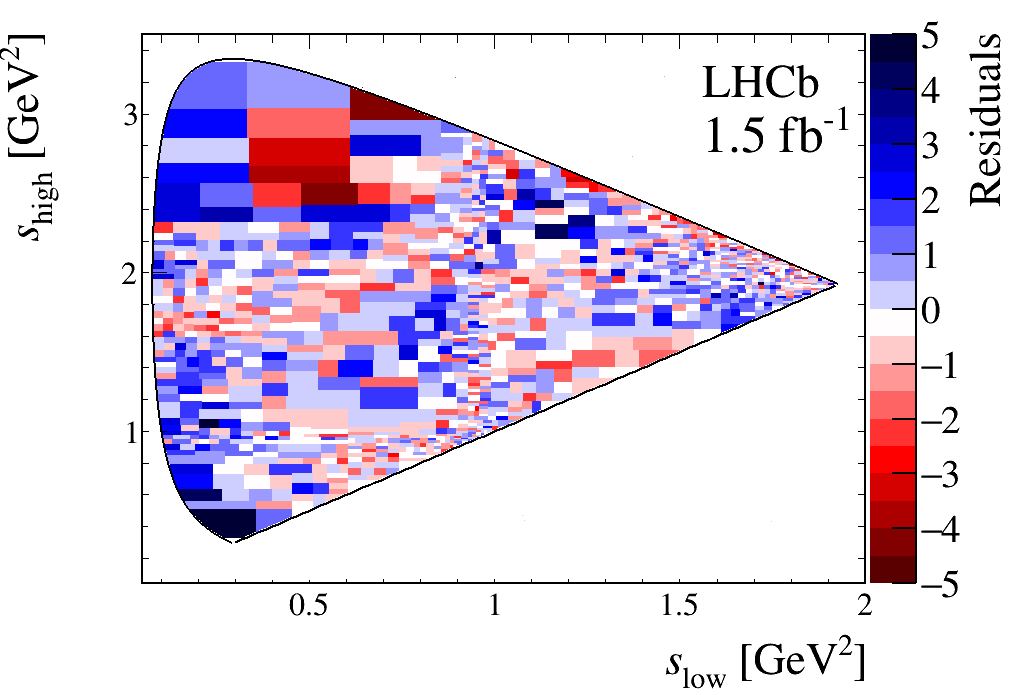}
    \caption{{(Left) The unfolded Dalitz plot projection $s_{13}$; (right) the folded distribution of the normalised residuals across the Dalitz plot.}}
    \label{fig:fitPWA1b}
\end{figure}
\begin{figure}[h!] 
 \centering
\includegraphics[width=0.49\textwidth]{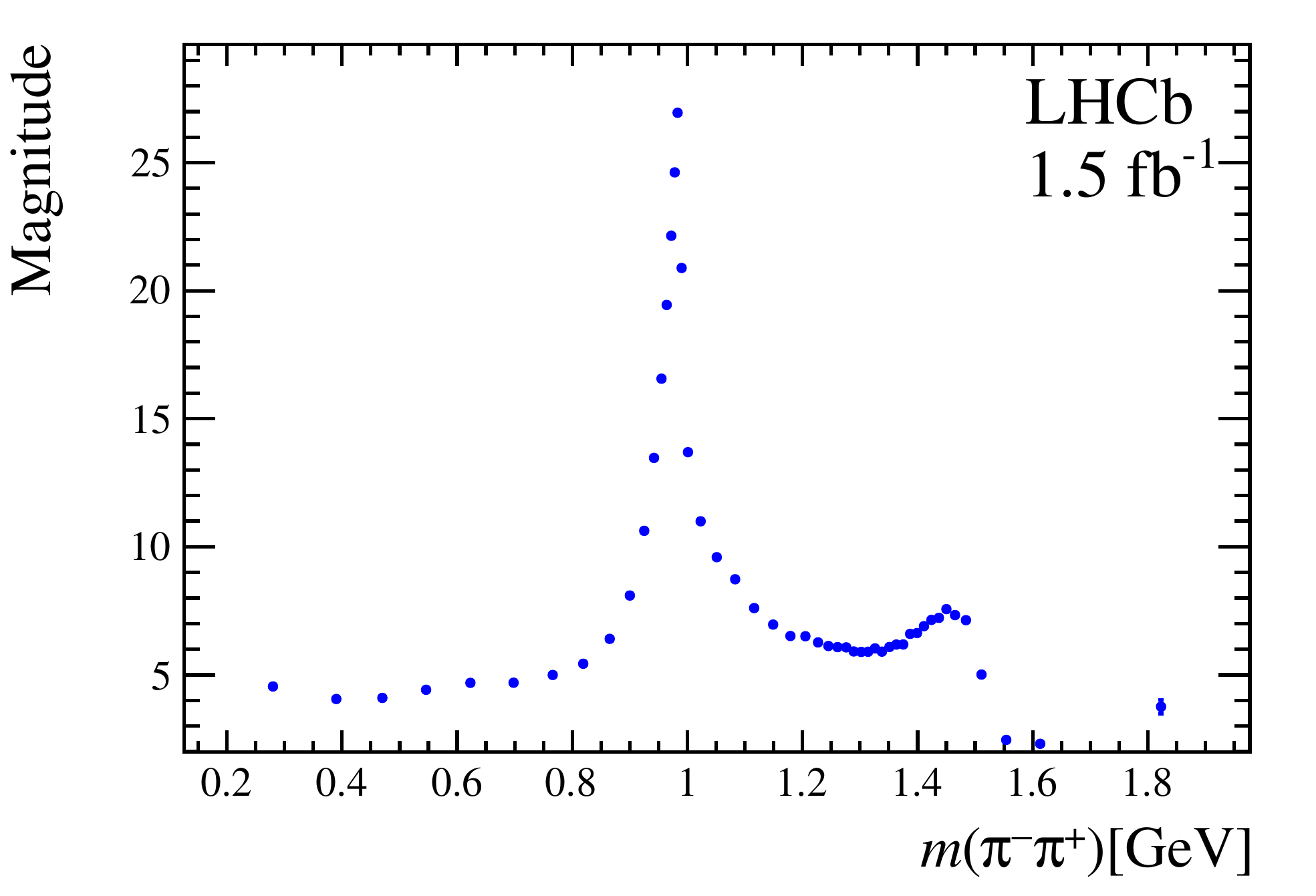}
\includegraphics[width=0.49\textwidth]{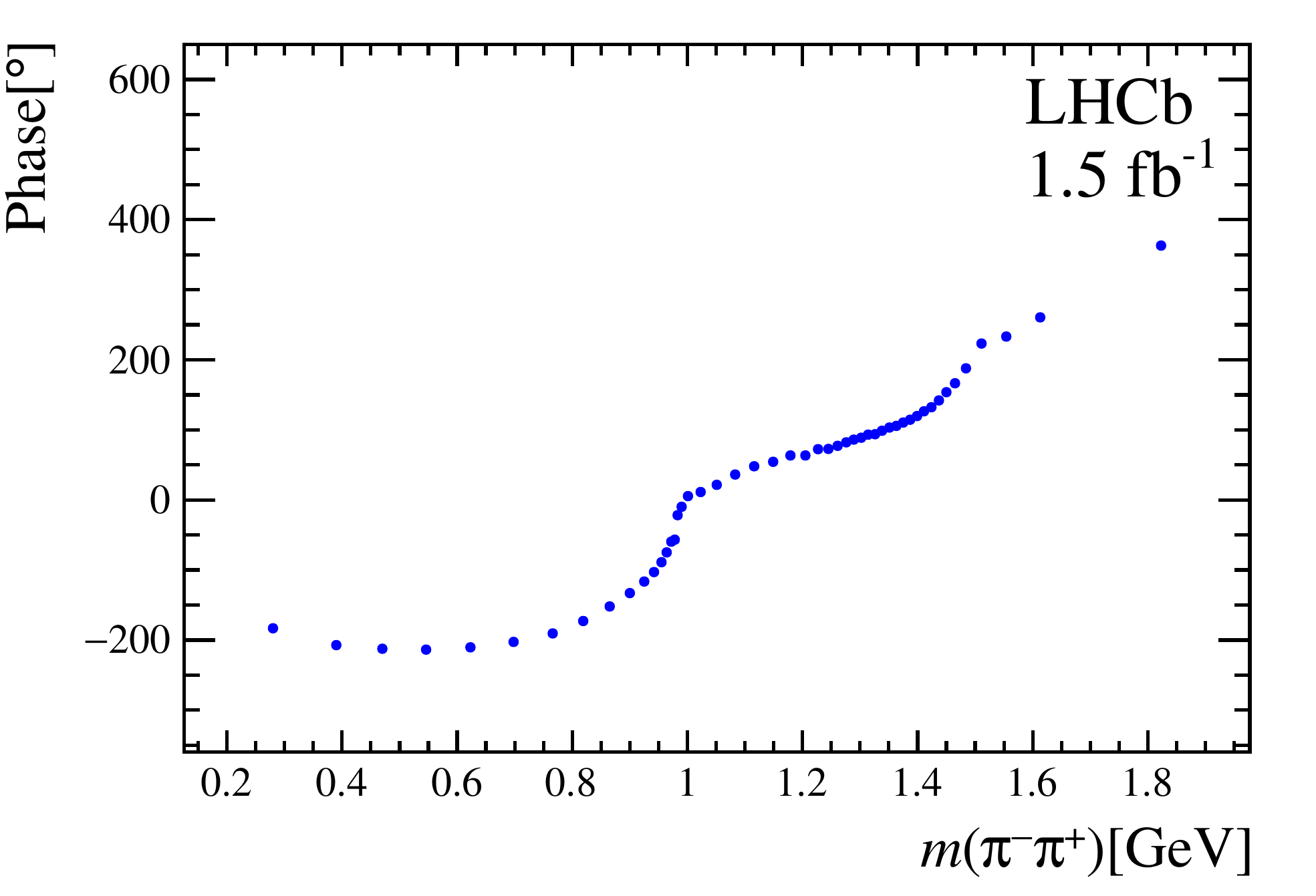}\caption{(Left) Magnitude  and (right) phase  of the  S-wave amplitude as a function of $m(\pi^+\pi^-)$. The uncertainties are statistical.}
\label{fig:SwavePWA1}

\end{figure}
\begin{figure}[hbt]
    \centering
    \includegraphics[width = 0.7\textwidth]{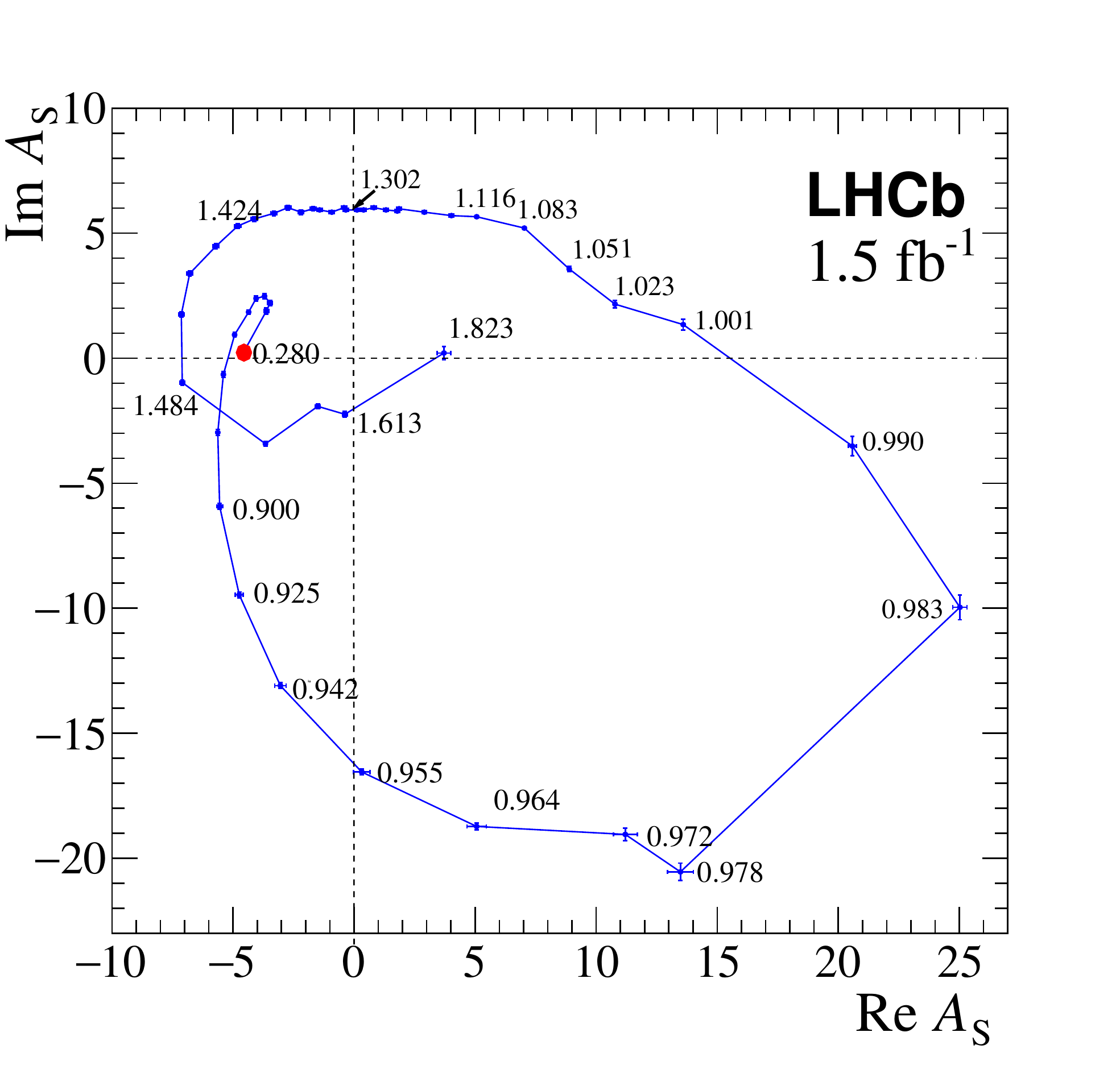}
    \caption{Argand plot of the $\pi^+\pi^-$ S-wave amplitude. The values of $m(\pi^+\pi^-)$ at the edge of each interval are indicated next to the corresponding experimental point. The amplitude starts at the point circled in red and undergoes two counterclockwise circles.}
    \label{fig:argand}
\end{figure}

The data projections of the highest ($s_{high}$) and the lowest ($s_{low}$) $\pi^{-}\pi^{+}$ mass are shown in Fig.~\ref{fig:fitPWA1a} with the fit result superimposed. Fig.~\ref{fig:fitPWA1b} shows the unfolded $\pi^{+}\pi^{+}$ mass ($s_{13}$) and the normalized residuals across the Dalitz plane. Due to the two identical pions in the final state, the normalised residuals are computed over the folded Dalitz plot.
The fit results, expressed in terms of magnitudes, relative phases and fit fractions of each resonant component, are summarised in Table~\ref{tab:fitPWA1}.  To allow a comparison with previous results, the magnitude and phase of the S-wave amplitude are displayed as a function of the $\pi^+\pi^-$ mass, $m(\pip\pim)$, and are presented in Fig.~\ref{fig:SwavePWA1}, with the corresponding Argand plot shown in Fig.~\ref{fig:argand}. The interference fit fractions are collected in Table~\ref{tab:PWA1_intFF}.

The main features of the default fit are:
\begin{itemize}
\item The resonant structure is dominated by the S-wave component, with a fit fraction of 85\%, in agreement with previous determinations~\cite{babar-ds,besiiicollaboration2021amplitude}, followed by the D-wave component and a small P-wave contribution. 

\item The magnitude of the S-wave component near threshold is small and nearly constant, indicating a very small contribution from the $f_0(500)$ meson, also known as $\sigma$. 

\item The large phase variation and the peak in the magnitude near $m(\pi^+\pi^-)\sim 1\gev$  are signatures of a prominent contribution of the $f_0(980)$ meson.

\item The rapid growth of the S-wave phase towards the end of the spectrum indicates the presence of at least one more scalar resonance, which could be the $f_0(1370)$ meson, the $f_0(1500)$ meson or a combination of both. 

\item The Argand plot exhibits two overlapping circles, supporting the existence of at least one scalar resonance above $1\gev$.

\item The $D^+_s\to \omega(782)\pi^+$, $D^+_s\to \rho(1700)^{0}\pi^+$ and $D^+_s\to f'_2(1525)\pi^+$ channels are observed for the first time in the \Dsppp decay.

\item There is a small contribution from the $\rho(770)^0\pi^+$ amplitude, in agreement with previous analyses and in contrast with the $D^+$ decay, where the contribution of this channel is $26.0\%$~\cite{LHCb-PAPER-2022-016}.

\item The combined fit fraction of the $\rho(1450)^0$ and $\rho(1700)^0$ amplitudes is six times larger than that of the $\rho(770)^0$ contribution.

\end{itemize}

\begin{table}[h!]
\centering
\caption{Results from the default fit. The row ``combined" has the combined fit fractions of the $\rho(1450)^0$ and $\rho(1700)^0$ contributions, including the interference between them. The uncertainties are statistical. }
\label{tab:fitPWA1}
\begin{tabular}{>{$}c<{$}| >{$}r<{\,\pm\,$}@{}>{$}l<{$} >{$}r<{\,\pm\,$}@{}>{$}l<{$} >{$}r<{\,\pm\,$}@{}>{$}l<{$}}
\toprule
\multicolumn{1}{c|}{Resonance} & 
\multicolumn{2}{c}{Magnitude} &
\multicolumn{2}{c}{Phase [$^\circ$]} & 
\multicolumn{2}{c}{Fit fraction (FF) [\%]} \\
\midrule
\multicolumn{1}{c|}{S-wave} & \multicolumn{2}{c}{} & \multicolumn{2}{c}{} & 84.97&0.14 \\
\rho(770)^{0}  & 0.1201	&0.0030	 & 79.4  &1.8 & 1.038&0.054 \\
\omega(782)    & 0.04001&0.00090 & -109.9&1.7 & 0.360&0.016 \\
\midrule
\rho(1450)^{0} & 1.277  &0.026	 & -115.2&2.6 & 3.86 &0.15  \\
\rho(1700)^{0} & 0.873  &0.061	 & -60.9 &6.1 & 0.365&0.050 \\
\multicolumn{1}{c|}{combined} & \multicolumn{2}{c}{--} & \multicolumn{2}{c}{--} & 6.14&0.27 \\
\midrule
f_2(1270) & \multicolumn{2}{c}{1 (fixed)} & \multicolumn{2}{c}{0 (fixed)} & 13.69&0.14 \\
f'_2(1525) & 0.1098&0.0069 & 178.1&4.2 & 0.0455&0.0070 \\ 
\midrule
\multicolumn{1}{c|}{sum of fit fractions} & \multicolumn{2}{c}{} & \multicolumn{2}{c}{} & \multicolumn{2}{c}{104.3} \\
\multicolumn{1}{c|}{$\chi^2$/ndof (range)} & \multicolumn{2}{c}{[$1.45 - 1.57$]} & \multicolumn{2}{c}{} & \multicolumn{2}{c}{}\\
\bottomrule
\end{tabular}
\end{table}

\begin{table} [h!]
\centering
\caption{Results from the default fit. Interference fit fractions (\%) between the resonant amplitudes. The uncertainties are statistical.\label{tab:PWA1_intFF}}
\newcommand{\dc}[1]{\multicolumn{2}{c}{#1}}
\begin{tabular}{>{$}c<{$}| >{$}r<{\,\pm\,$}@{}>{$}l<{$} >{$}r<{\,\pm\,$}@{}>{$}l<{$} >{$}r<{\,\pm\,$}@{}>{$}l<{$} >{$}r<{\,\pm\,$}@{}>{$}l<{$}}
\toprule
& \dc{$\omega(782)$} & \dc{$\rho(770)^{0}$} & \dc{$\rho(1450)^{0}$} &  \dc{$\rho(1700)^{0}$} \\
\midrule
\omega(782)                  &   0.360&0.016   & \dc{} & \dc{} & \dc{} \\
\rho(770)^{0}                &   0.128&0.013   &   1.038&0.054  & \dc{} & \dc{} \\
\rho(1450)^{0}               &    0.36&0.14    &   0.148&0.14   &   3.86&0.15   & \dc{} \\
\rho(1700)^{0}               &   0.089&0.010   &  -0.307&0.0.55 &   1.92&0.20   &  0.365&0.050  \\
f_2(1270)                    & -0.1540&0.0040  &   0.280&0.029  &  -1.10&0.047  & -0.376&0.047  \\
f'_2(1525)                   & 0.00827&0.00063 & 0.00283&0.0038 &  0.066&0.0021 & 0.0200&0.0021 \\
\multicolumn{1}{c|}{S-wave} &  -0.053&0.0099  &   0.804&0.076  & -1.520&0.086 & -0.934&0.086  \\
\midrule[\heavyrulewidth]
& \dc{$f_2(1270)$} & \dc{$f'_2(1525)$} & \dc{S-wave} \\
\midrule
f_2(1270)                    &  13.69&0.14  & \dc{} & \dc{} \\
f'_2(1525)                   & -0.429&0.072 & 0.0455&0.0070 & \dc{} \\
\multicolumn{1}{c|}{S-wave} & -3.460&0.092 &   0.20&0.013  & 84.97&0.14 \\
\bottomrule
\end{tabular}

\end{table}


\section{Systematic uncertainties}
\label{sec:systematics}

The systematic uncertainties are divided into two types: those related to experimental effects  and the ones associated with the parameters of the model. The experimental systematic uncertainties account for the uncertainties on the efficiency, on the background model, on the mass resolution and a possible bias from the fitting algorithm. 

The systematic uncertainties on the efficiency include uncertainties on PID efficiencies, the impact of the binning choice of the efficiency histogram before the 2D spline smoothing and the uncertainties due to the limited size of the simulation sample. The uncertainties on the PID efficiency are related to the finite size and the binning scheme of the calibration samples.  Alternative sets of PID efficiency weights are produced by varying the binning scheme and the candidate efficiency according to the statistical uncertainties of the calibration samples. These alternative sets of PID weights are applied to the simulation sample and the data are fitted using the resulting signal efficiency histograms. The root mean square of of the resulting distribution of each fit parameter is assigned as the systematic uncertainty. The systematic uncertainty due to the binning scheme of the efficiency histogram is assessed by varying the number of bins ($20\times20$ and $25\times25$ bins), and for each fit parameter the largest deviation is assigned as the systematic uncertainty. The impact of the limited size of the simulation sample is estimated by fitting the data with 100 alternative efficiency histograms, in which the bin contents are varied according to a Poisson distribution. The root mean square of the resulting distribution of each fit parameter is assigned as the systematic uncertainty on each fit parameter. The systematic uncertainty on the efficiency is the sum in quadrature of the various sources considered. 
 
The uncertainties on the background model include the effect of the histogram binning before the 2D-spline smoothing, the weights assigned to each sideband and the uncertainty on the signal to background ratio from the $\pi^-\pi^+\pi^+$ invariant-mass fit. Alternative background histograms are produced using $20\times20$ and $25\times25$ bins, and for each fit parameter the largest deviation from its default value is assigned as a systematic uncertainty. The data are fitted with the weight of each sideband varied from 0\% to 100\% and the largest deviation is assigned as a systematic uncertainty. The signal to background ratio is varied according to the uncertainty of the invariant-mass fit, and the largest deviation is assigned as a systematic uncertainty. All the above uncertainties are added in quadrature and the result is assigned as the systematic uncertainty on the background.

The default fit is obtained with a convolution of the Breit--Wigner representing the $\omega(782)$ line shape with the mass resolution function. A systematic uncertainty is assigned by varying the width of the resolution function according to its uncertainty. The differences with respect to the default fit are taken as systematic uncertainties.

Systematic uncertainties arising from biases in the fit algorithm are estimated using an ensemble of 1000 pseudoexperiments generated according to the fitted values of the parameters. The simulations include the background component and the effect of the efficiency variation across the Dalitz plot. Each pseudoexperiment is fitted independently, resulting in a distribution of fitted values for each parameter. The difference between the mean of the distributions and the default value of each parameter is assigned as the systematic uncertainty due to the fit bias.

The systematic uncertainty due to the decay amplitude model includes the uncertainties on the masses and widths of the resonances, the choice of the number of intervals in the model-independent description of the S-wave amplitude and the uncertainty on values of the Blatt--Weisskopf parameters $r_R$ and $r_D$. The masses and widths of the resonances are always fixed in the fits, but their values are varied individually, according to their uncertainty quoted in the PDG averages~\cite{PDG2022}. The largest deviations of the fit parameters are taken as systematic uncertainties. The fit is repeated with the default model, dividing the $\pi^+\pi^-$ mass squared spectrum into 45 and 55 intervals. As before, the largest deviation is assigned as systematic uncertainty on each parameter. Finally, the largest difference from the default value of each fit parameter is assigned as a systematic uncertainty when the values of the Blatt--Weisskopf parameters are varied in the range 1.0--2.0 $\gev^{-1}$, for $r_R$, and 4.0--6.0$\gev^{-1}$, for $r_D$. The systematic uncertainties due to the decay amplitude model is the sum in quadrature of the above contributions.

The systematic uncertainties on the magnitudes, phases and fit fractions of P- and D-wave resonances are summarised in Tables~\ref{tab:total_syst} and \ref{tab:ffs}. The statistical uncertainties are quoted in the last columns, showing that the measurement is dominated by the systematic uncertainty. For most parameters, the background model is the dominant source of experimental systematic uncertainty. For almost all amplitudes, the systematic uncertainty due to the decay amplitude model is the dominant source.

\begin{table}[h!]
\caption{Summary of the systematic uncertainties on the magnitudes and phases of the complex coefficients from the P- and D-wave amplitudes. The column ``total exp." is the sum in quadrature of the first four columns. For comparison, the statistical uncertainties are included in the last column.\label{tab:total_syst}}
\centering
\begin{tabular}{l |c c c c c c c}
\toprule
parameter  &  back.  &  eff.  & fit bias & mass res. & total exp.  &   model &  stat. \\
\midrule
$\rho(770)^0$ mag.	&	0.0034	&	0.0030	&	0.0021	&	0.0000	&	0.0050	&	0.0062	&	0.0030	\\
$\rho(770)^0$ ph.	&	6.1	&	4.6	&	1.1	&	0.016	&	7.8	&	4.4	&	1.8	\\
$\omega(782)$ mag.	&	0.0017	&	0.00047	&	0.00041	&	0.00003	&	0.0018	&	0.00086	&	0.00090	\\
$\omega(782)$ ph.	&	0.71	&	0.53	&	0.32	&	0.0030	&	0.94	&	1.4	&	1.7	\\
$\rho(1450)^0$ mag.	&	0.011	&	0.016	&	0.013	&	0.000	&	0.023	&	0.48	&	0.026	\\
$\rho(1450)^0$ ph.	&	1.2	&	0.72	&	2.5	&	0.011	&	2.83	&	10.4	&	2.6	\\
$\rho(1700)^0$ mag.	&	0.041	&	0.032	&	0.014	&	0.000	&	0.054	&	0.62	&	0.061	\\
$\rho(1700)^0$ ph.	&	3.1	&	3.5	&	4.7	&	0.017	&	6.7	&	12	&	6.1	\\
$f'_{2}(1525)$ mag.	&	0.0080	&	0.017	&	0.0031	&	0.0000	&	0.019	&	0.015	&	0.0069	\\
$f'_{2}(1525)$  ph.	&	8.5	&	7.9	&	1.2	&	0.012	&	12	&	7.0	&	4.2	\\
\bottomrule
\end{tabular}
\end{table}

\begin{table}[h!]
\caption{Systematic uncertainties on the fit fractions (\%).  The combined fit fractions of the $\rho(1450)^{0}\pi^+$ and $\rho(1700)^{0}\pi^+$ are quoted in the row ``combined".\label{tab:ffs}}
\centering
\begin{tabular}{l |c  c  c  c  c  c}
\toprule
amplitude         & back.  & eff.      & fit bias  & total exp.&   model   &  stat. \\
\midrule
$S$-wave	          &	0.28	&	0.085	&	0.082	&	0.30	&	0.63	&	0.14	\\
$\rho(770)^0\pi^+$	  &	0.066	&	0.061	&	0.037	&	0.097	&	0.11	&	0.054	\\
$\omega(782)\pi^+$	  &	0.031	&	0.013	&	0.0072	&	0.034	&	0.016	&	0.016	\\
\midrule
$\rho(1450)^{0}\pi^+$ &	0.041	&	0.10	&	0.078	&	0.14	&	2.0	    &	0.15	\\
$\rho(1700)^{0}\pi^+$ &	0.035	&	0.027	&	0.0095	&	0.045	&	0.34	&	0.050	\\
combined  &	0.10	&	0.20	&	0.26	&	0.34	&	1.9	&	0.27	\\
\midrule
$f_2(1270)\pi^+$      &	0.12	&	0.19	&	0.0026	&	0.22	&	0.49	&	0.14	\\
$f'_2(1525)\pi^+$     &	0.007	&	0.013	&	0.0029	&	0.015	&	0.0087	&	0.0070	\\
\midrule
\end{tabular}
\end{table}

\section{The S-wave: comparison with $\boldsymbol{\Dppp}$ decay}
\label{sec:DpDs}

\begin{figure}[h] 
 \centering
\includegraphics[width= 0.48\textwidth]{figs/Fig3a.pdf}
\includegraphics[width= 0.48\textwidth]{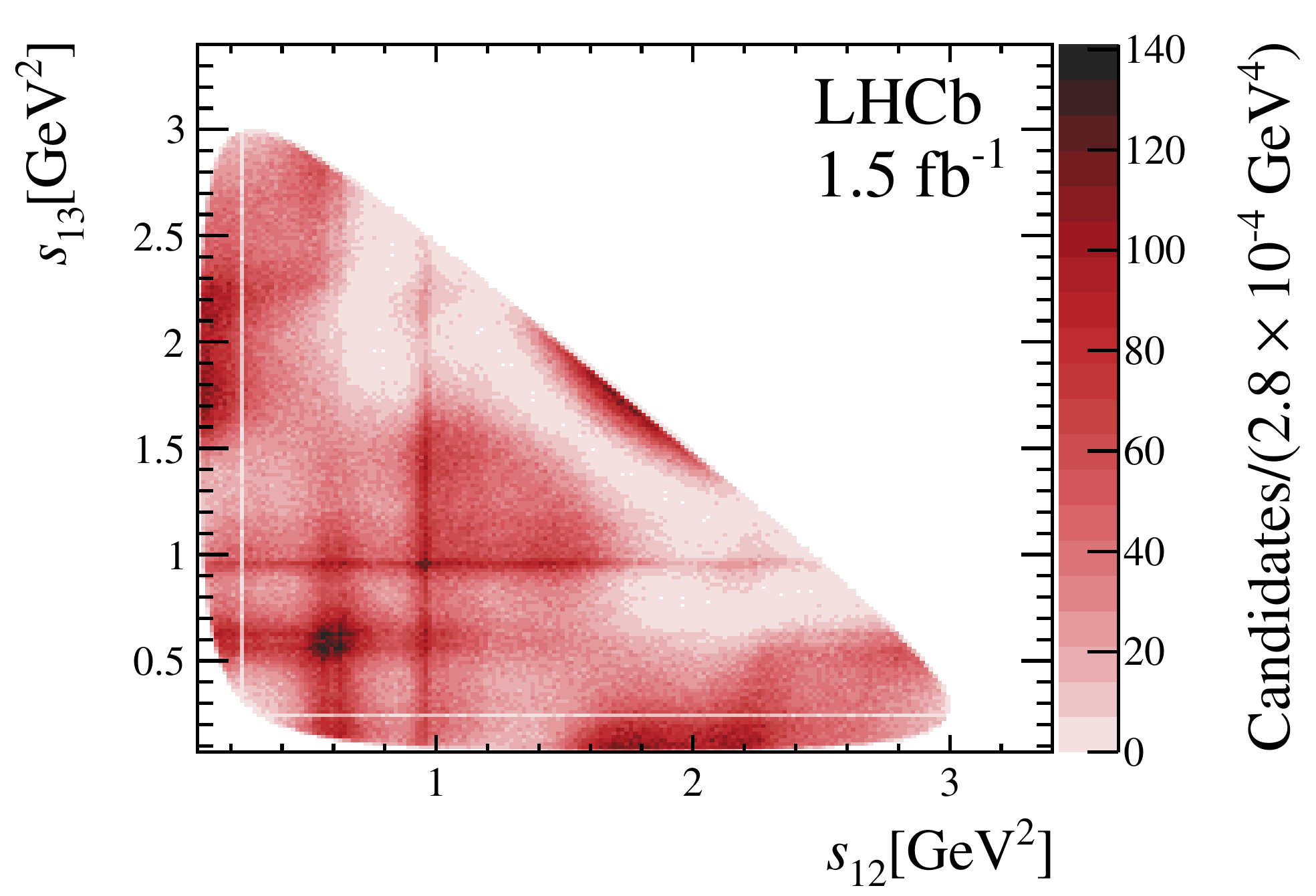}
\caption{(Left) Dalitz plot of the \Dsppp  and (right) \Dppp decays~\cite{LHCb-PAPER-2022-016}. The colour scale indicates the density of candidates.}
\label{fig:Dps}
\end{figure}

\begin{figure}[h] 
 \centering
\includegraphics[width=0.7\textwidth]{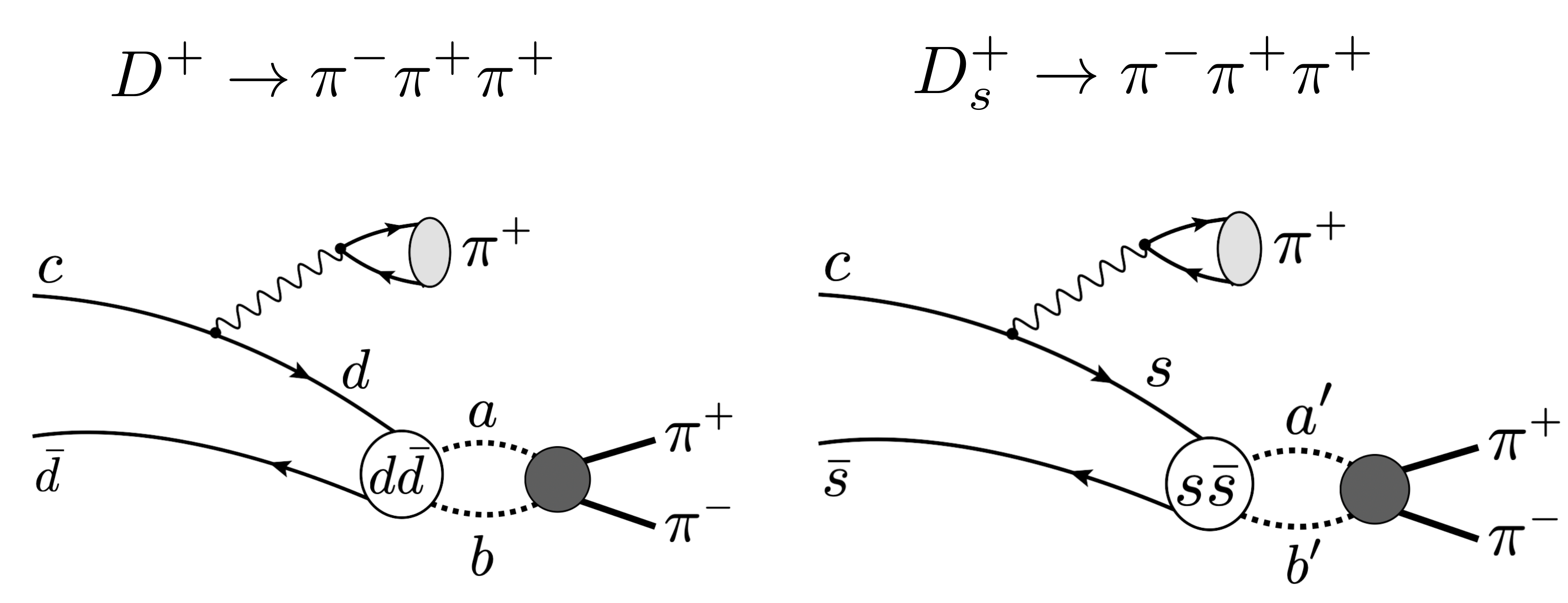}

\caption{Dominant amplitudes for the \Dppp and \Dsppp decays. The scalar resonances are produced from rescattering of the two pseudoscalar mesons $ab$ ($a'b'$) formed by a $q\bar q$ pair from the vacuum and the $d\bar d$ ($s\bar s$) pair from the decay of the $c$ quark. }
\label{fig:unit-qm}
\end{figure}

The Dalitz plots of the \Dsppp and \Dppp decays~\cite{LHCb-PAPER-2022-016}, shown in Fig.~\ref{fig:Dps}, reveal different resonant structures. It is widely accepted that the tree-level $W$-emission amplitude, illustrated in Fig.~\ref{fig:unit-qm}, is the dominant mechanism in $D$-meson decays. The resonances would be produced from an $s\bar s$ source in \Dsppp decays, and from a $d\bar d$ source in \Dppp decays.  Similar differences in the resonant structure are observed in the $B^0_{(s)}\to\jpsi\pi^+\pi^-$ decays~\cite{LHCb-PAPER-2014-012,LHCB-PAPER-2013-069}. In these decays, the \jpsi recoils against a $d\bar d$ pair, in the case of the $B^0$ decay, and against an $s\bar s$ pair, in the case of the $B^0_s$ decay. To a very good approximation, the interaction between the \jpsi and the $\pi^+\pi^-$ system can be ignored. The production of the $\pi^+\pi^-$ system from  $s\bar s$ and $d\bar d$ sources results in different resonant structures.

\begin{figure}[h] 
 \centering
\includegraphics[width= 0.49\textwidth]{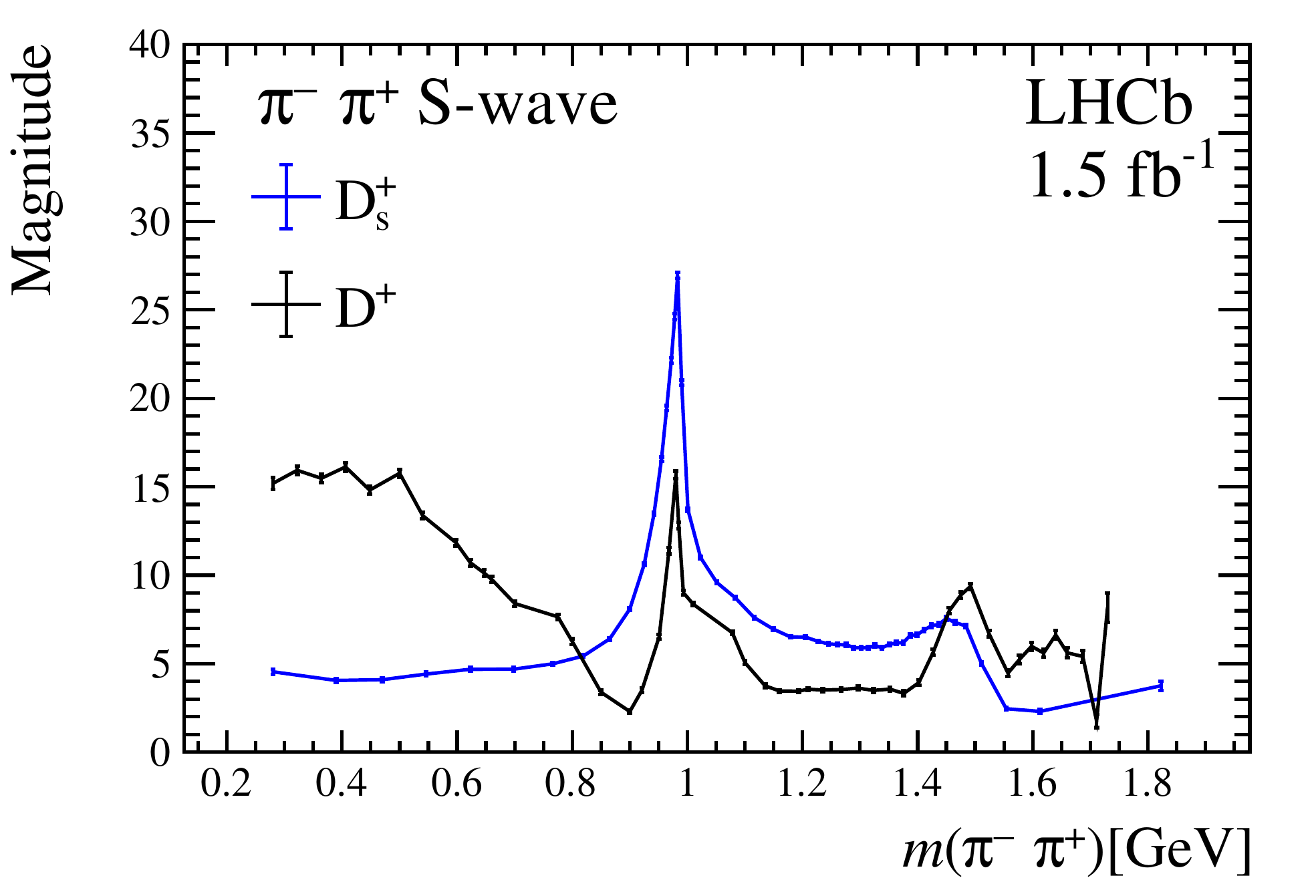}
\includegraphics[width= 0.49\textwidth]{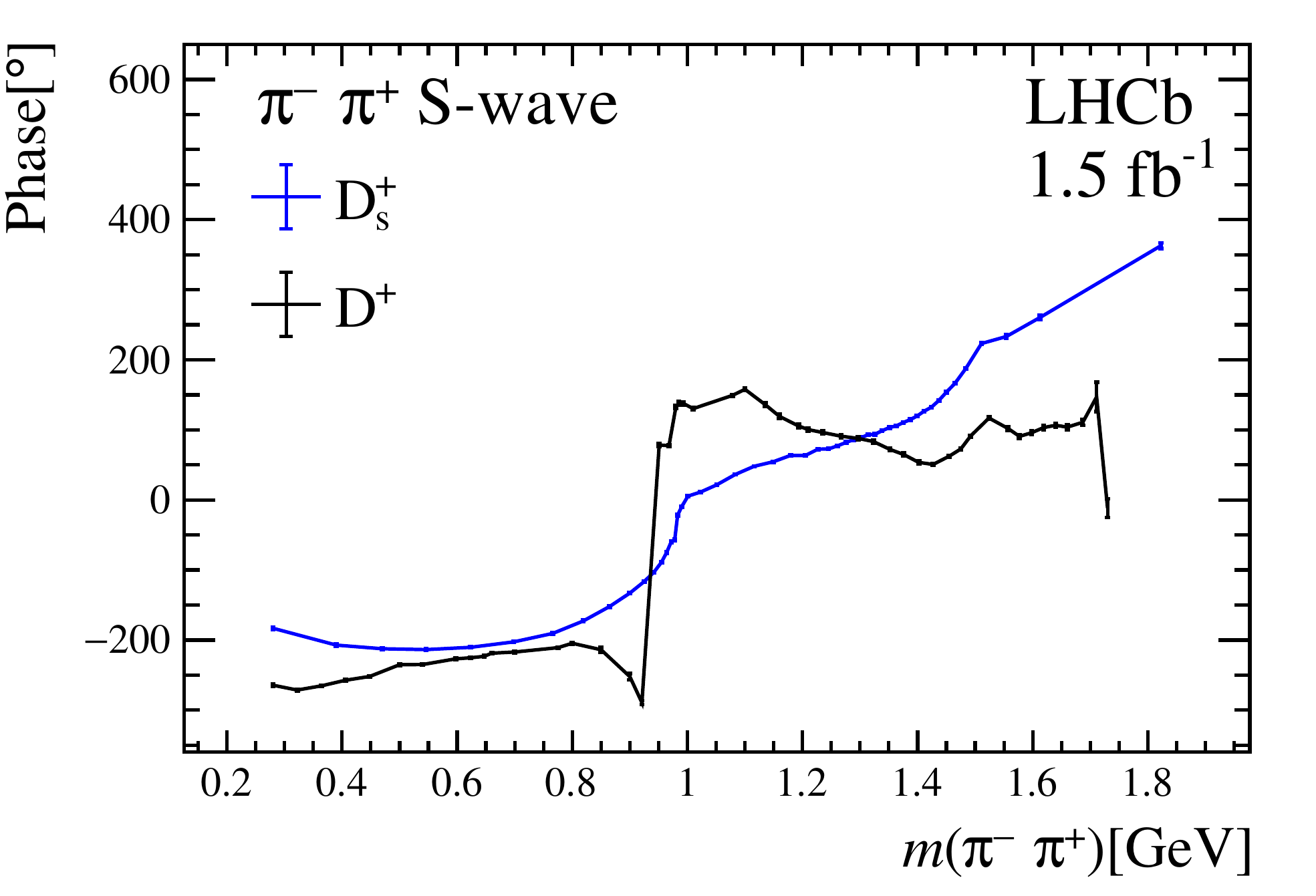}
\caption{(Left) Magnitude and (right) phase of the \pim\pip~S-wave amplitude for the \mbox{\Dppp } (black line) and \Dsppp~decays (blue dot). }
\label{fig:DDsSW}
\end{figure}

A comparison between the S-wave magnitude and phase from the \Dsppp and \Dppp decays is presented in Fig.~\ref{fig:DDsSW}. 
A broad structure in the magnitude of the \Dppp S-wave is observed in the beginning of the $\pi^+\pi^-$ spectrum but it is absent in the \Dsp decay. Model-dependent analyses~\cite{CleoD3pi} attribute this structure to the resonant mode $f_0(500)\pi^+$. This channel accounts for nearly half the \Dppp decay rate, but is consistent with zero in the \Dsppp decay~\cite{babar-ds}. In the latter, the $f_0(980)\pi^+$ mode is dominant, corresponding to approximately half the decay rate.

 The vector and tensor resonances match the $q\bar q$ states predicted by the quark model, so they could therefore couple directly to the $D$ meson. However, the overpopulation of scalar states below $2\gev$ suggests that at least some of these resonances may not be regular $q\bar q$ states. The $f_0(980)$ and the $a_0(980)$ states, for instance, are often interpreted as compact tetraquark states, see 
``Note on Scalar Mesons Below $2\gev$"  in Ref.~\cite{PDG2020}, while the $f_0(500)$ and the $K^*(800)$ states could be dynamically generated poles of $\pi\pi\to\pi\pi$ and $K\pi\to K\pi$ scattering~\cite{paper:kaminski2008pion}, respectively. 

The differences between the $\pi^+\pi^-$ S-wave amplitudes in \Dsppp and \mbox{\Dppp} decays may be understood in the framework of the unitary chiral model~\cite{paper:oset}. The production of a pair of pseudoscalar mesons with zero orbital angular momentum could be energetically favoured compared to a scalar particle, \eg quarks with spins aligned in an $L=1$ state. The scalar mesons would be produced by the rescattering of the pair of pseudoscalar particles $ab\to\pi^+\pi^-$ ($a,b=\pi, K,\eta$), as illustrated in Fig.~\ref{fig:unit-qm}. In this picture, the $d\bar d$ and $s\bar s$ pairs combine with $q\bar q$ pairs from the vacuum ($q=u,\ d,\ s$), giving rise to different sets of pseudoscalar mesons and, therefore, to different S-wave amplitudes.

Considering the three possible light-quark pairs from the vacuum inserted between the $d\bar d$ pair, the $D^+$ decay has 
\begin{equation}
d(\bar u u+\bar d d+\bar s s)\bar d = d\bar u u\bar d + d\bar d d\bar d + d\bar s s\bar d~,
\end{equation}
 which, in terms of the pseudoscalar mesons, corresponds to 
\begin{equation}
\sum_i d\bar q_i q_i\bar d = \pi^+\pi^-+\tfrac{1}{2}\pi^0\pi^0- \tfrac{2}{\sqrt{6}}\pi^0\eta+K^0\ensuremath{\Kbar{}^0}\xspace + \tfrac{1}{3}
\eta\eta~.
\end{equation}

In the $D^+\to \pi^-\pi^+\pi^+$ decay, the S-wave would be formed by the reactions $ab\to \pi^+\pi^-$, with $ab=\pi^+\pi^-,\pi^0\pi^0, \pi^0\eta, K^0\ensuremath{\Kbar{}^0}\xspace$ and $\eta\eta$. In the $D^+_s\to \pi^-\pi^+\pi^+$ decay, $s(\bar u u+\bar d d+\bar s s)\bar s$ corresponds to 
\begin{equation}
\sum_i s\bar q_i q_i\bar s  = K^+K^-+K^0\ensuremath{\Kbar{}^0}\xspace + \tfrac{1}{3}\eta\eta~,
\end{equation}
leading to a different set of reactions  $a'b'\to \pi^-\pi^+$, and therefore to different S-wave amplitudes. In this picture, the lack of an $f_0(500)$ contribution in the $D^+_s$ decay supports to the interpretation of this resonance as a dynamical pole of the $\pi\pi$ scattering. The $f_0(980)$ resonance is known to couple strongly to $K\Kb$, which can explain the relative prominence of this state in $D^+_s$ decays with respect to $D^+$ decays.

\section{The S-wave: comparison with $\boldsymbol{\pi^+\pi^-}$ scattering}
\label{sec:Dsvspipi}
\begin{figure}[h] 
\centering
\includegraphics[width=0.49\textwidth]{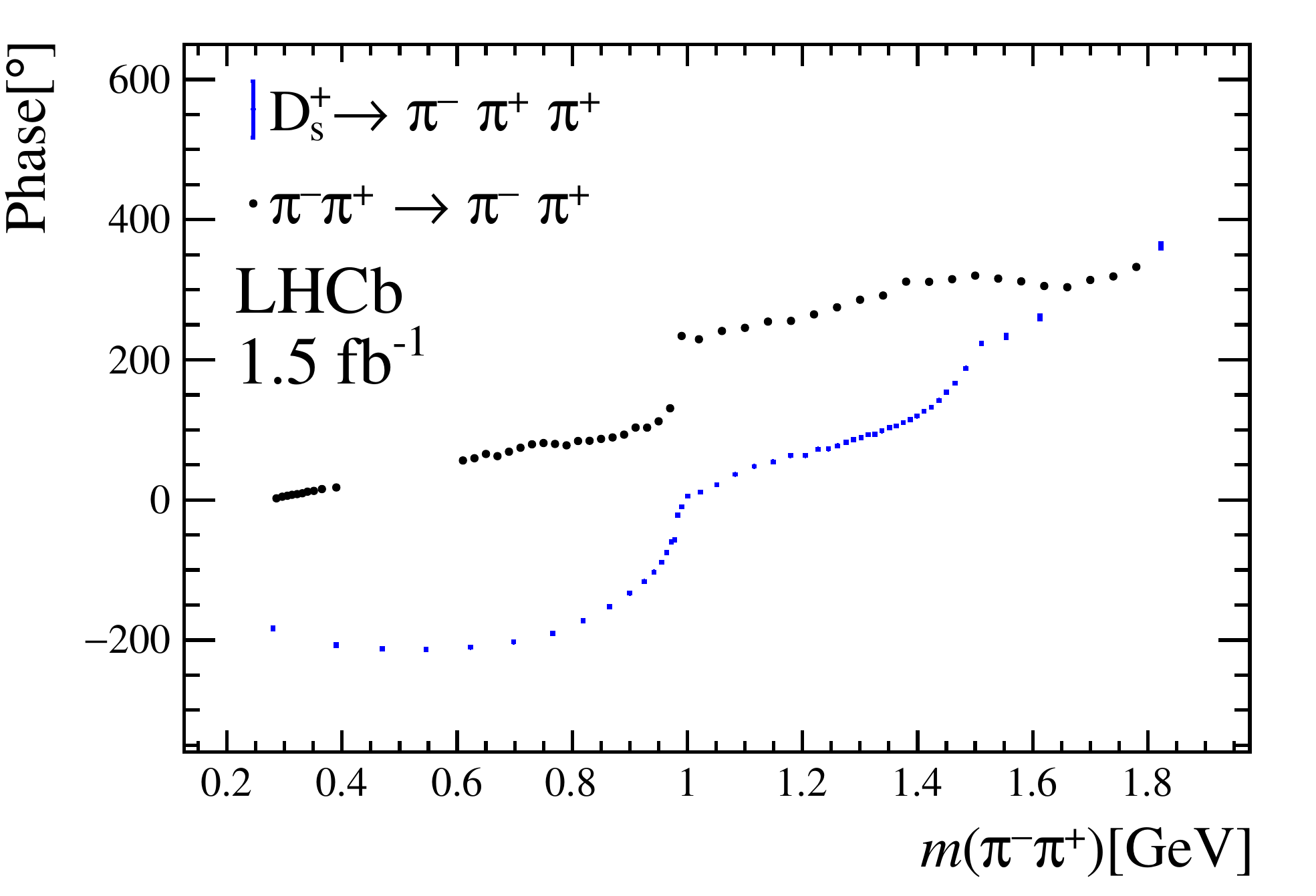}\includegraphics[width=0.49\textwidth]{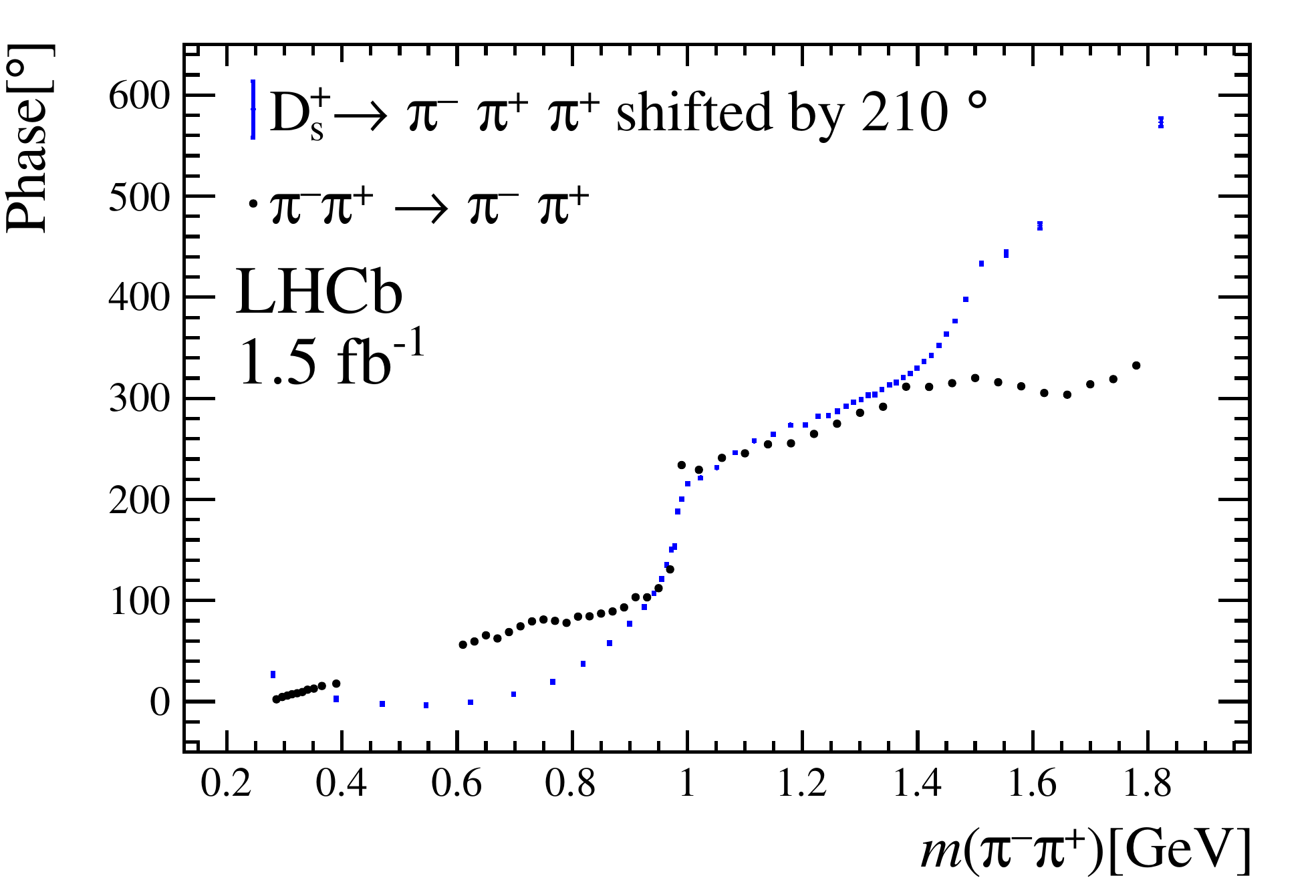}
\caption{(Left) Comparison of the $\pi^+\pi^-$ S-wave phase from \Dsppp decays and the phase from $\pi^+\pi^-\to \pi^+\pi^-$ scattering.  Data on $\pi^+\pi^-\to \pi^+\pi^-$ scattering above 0.6\gev are from a re-analysis~\cite{Ochs_2013} of original data from~\cite{hyams}, and below 0.4\gev  are from $K_{e4}$ decays~\cite{NA48}. (Right) The S-wave phase from the \Dsppp decay is shifted by 210$^{\circ}$.}
\label{fig:DsScatt}
\end{figure}

In Figure~\ref{fig:DsScatt}, the phase of the S-wave amplitude from the \Dsppp decay is compared to the scalar-isoscalar phase shift $\delta^0_0$ from $\pi^+\pi^-\to \pi^+\pi^-$ scattering~\cite{Ochs_2013}. Significant differences between the two phases are observed. In $\pi^+\pi^-\to \pi^+\pi^-$ scattering, the phase starts from zero at threshold, as required by chiral symmetry. In the \Dsppp  decay, the phase starts at approximately $-200^\circ$, and this overall shift could be attributed to the production of the $\pi^+\pi^-$ pair. The differences in the shape of the two phases are more evident when the phase from the $D^+_s$ decay is shifted by 210$^{\circ}$. The $\pi\pi$ scattering is elastic up to the threshold of the $K\Kb$ channel. Near 1\gev, where the phase is dominated by the $f_0(980)$ resonance, the shape of the two phases is in qualitative agreement, but below 1\gev, where the $\pi\pi$ scattering is elastic, the two phases are clearly incompatible. In the inelastic regime, the two phases can no longer be directly compared. Nevertheless, 

\noindent
in both cases an acceleration of the phase motion at higher values of $m(\pi^+\pi^-)$ is observed, indicating the presence of one or more scalar resonances. 

Meson-meson interactions have a universal character, and this is the essence of Watson's theorem~\cite{watson1952effect}. If this theorem were to hold in $D$ meson decays, the phases $\delta^0_0$ and those of the S-wave amplitude from the \Dsppp and \Dppp decays should be the same. However, differences between these three phases are observed. 

The primary source of scattering data is the reaction $\pi N\to \pi\pi N$ (or $KN\to K\pi N$), which, at low momentum transfer, is dominated by one-pion exchange. The virtual pion is assumed to be nearly on-mass-shell, yielding the reaction $\pi^+\pi^-\to \pi^+\pi^-$, where the outgoing $\pi^+\pi^-$ pair recoils against the nucleon. In the \Dsppp decay, the $\pi^+\pi^-$ pair is produced in a quite different environment. The $\pi^+\pi^-$ pair is part of a three-body system, and may be produced from processes such as $D_s^+\to K^+K^-\pi^+\to \pi^-\pi^+\pi^+$. In the scenario where the scalar resonances arise from interactions of two pseudoscalar particles, the phase of the S-wave amplitude obtained from \Dsppp decays results from a set of coupled channels, $ab\to\pi^+\pi^-$ ($a,b=\pi,K$ or $\eta$), in contrast with the $\delta^0_0$ phase, which is obtained from a single reaction.

\section{The P- and D-waves}
\label{sec:PandD}
A distinct feature of the $\pi^+\pi^-$ P-wave amplitude in the \Dsppp decay is the small contribution of the $\rho(770)^0\pi^+$ channel, with a fit fraction of ($1.038 \pm 0.050$)\%, compared to ($26.0 \pm 0.3$)\% measured in the \Dppp decay. The small $\rho(770)^0$ component in the \Dsppp decay occurs because resonances with no strange quarks in the wave function can only be formed through the suppressed $W$-annihilation amplitude.  

In the P-wave amplitude, there is also a small contribution of $(0.360 \pm 0.016)$\% from the $D^+_s\to\omega(782)\pi^+$ channel. The $\omega(782)\to \pi^+\pi^-$ decay is  isospin violating, with a branching fraction of ($1.53^{+0.11}_{-0.13}$)\%~\cite{PDG2022}. Due to the small difference between the masses of the $u$ and $d$ quarks, isospin symmetry is only approximate. The physical $\rho(770)^0$ and $\omega(782)$ resonances are linear combinations of the pure isovector and isoscalar SU(3) states $|\rho_I\rangle$ and $|\omega_I\rangle$~\cite{gourdin},
\begin{equation}
|\rho(770)^0\rangle = |\rho_I\rangle -\epsilon_{\rho\omega} |\omega_I\rangle,\hskip .4cm    |\omega(782)\rangle = \epsilon_{\rho\omega}|\rho_I\rangle + |\omega_I\rangle~, 
\end{equation}
where $\epsilon_{\rho\omega}$ is a complex parameter associated with the mixing between the physical states.

The $\omega(782)\to \pi^+\pi^-$ decay has been observed in different reactions ~\cite{CBarrel-rw,CMD,LHCb-PAPER-2014-012,LHCb-PAPER-2019-017,LHCb-PAPER-2021-045,LHCb-PAPER-2022-016,KLOE:2003kas}, always in conjunction with a prominent $\rho(770)^0$ signal. In the \Dppp and \mbox{\Dsppp} decays, the $\omega(782)$ signal may arise either through the $\rho$-$\omega$ mixing or from the direct transition. In this work, the latter mechanism is assumed in the default fit, and is represented by a coherent sum of individual $\rho(770)^0\pi^+$ and $\omega(782)\pi^+$ amplitudes.
An alternative fit was performed replacing the individual amplitudes by the $\rho$--$\omega$ mixing amplitude used in Ref.~\cite{LHCb-PAPER-2019-017}, which ignores the direct $\omega(782)\to \pi^-\pi^+$ transition. The alternative and default fits have similar quality and yield the same S- and D-wave amplitudes. The $\rho$--$\omega$ fit fraction is the same as that of the combined contributions from the individual amplitudes in the default fit. The line shape and phase motion of the two parameterisations are nearly identical, preventing the separation of the direct production $\omega(782)$ from the transition through $\rho$--$\omega$ mixing.

In the \Dppp decay~\cite{LHCb-PAPER-2022-016}, where the production of the $\rho(770)^0$ is favoured, the ratio between the $\rho(770)^0$ and $\omega(782)$ fit fractions is approximately 250. A different scenario is observed in the \Dsppp decay, where the contribution from the $\rho(770)^0$ is only $\sim1\%$ and the ratio between the $\omega(782)$ and $\rho(770)^0$ fit fractions is 3.1, suggesting different mechanisms for the production of these resonances between both decays.

An alternative mechanism is proposed in Refs.~\cite{cheng} and~\cite{yu}. The $\omega(782)$ meson  could also be produced by the final-state rescattering $D^+_s\to \eta(\eta^{'})\rho(770)^+ \to \omega(782)\pi^+$. Due to $G$-parity conservation, the rescattering $\eta(\eta^{'})\rho(770)^+\to \rho(770)^0\pi^+$ cannot occur. Amplitudes of this type involve quantum loops that are suppressed, but there is a compensation from the large branching fractions $\mathcal{B}(D^+_s\to \eta\rho(770)^+) = (8.9\pm 0.8)\%$ and $\mathcal{B}(D^+_s\to \eta'\rho(770)^+) = (5.8\pm 1.5)\%$~\cite{PDG2022}. This could explain why the fit fraction of the $\omega(782)\pi^+$ amplitude is three times larger in the \Dsppp decay than in \Dppp decay.

The most significant components of the P-wave amplitude are the $\rho(1450)^0\pi^+$ and $\rho(1700)^0\pi^+$ channels, with similar contributions to the \Dppp and \Dsppp decays. There are a number of measurements, \eg $\bar pd$~\cite{cbarrel1} and $p\bar p$~\cite{obelix} annihilation at rest, $\tau^- \to \pi^-\pi^0\nu_{\tau}$ decays~\cite{paper:belle-tau} and $e^+e^-\to\pi^+\pi^-\gamma$ cross-section with initial state radiation~\cite{paper:babar-isr}, whose description requires two interfering vector resonances in the region $1.4\lesssim m(\pi^+\pi^-)\lesssim1.8\gev$.  

The determination of the individual contributions of the $\rho(1450)^0$ and $\rho(1700)^0$ is limited by a strong correlation between the fit fractions and the masses and widths of these states. However, the combined fit fraction remains approximately constant when different values of the masses and widths of theses states are used in the fit. Additionally, the uncertainties on the masses and widths have a small impact on the S- and D-wave amplitudes. The combined fit fractions are ($6.14\pm 0.27$)\%, in the \Dsppp decay, and ($7.1 \pm 0.9$)\% in the \Dppp decay, indicating that these resonances have a significant $s\bar s$ component in their wave functions. This is consistent with what is observed in $\pi\pi$ scattering, where the inelasticity between 1 and $1.8\gev$ is mostly due to the threshold of the $K$\Kb channel~\cite{hyams}. The analysis of $\pi\pi$ scattering data~\cite{pelaez19} reveals that the inelasticity of the P-wave is close to one up to $1.4\gev$, after which it decreases to less than 0.5 at $\sim1.6\gev$, returning to one at $1.8\gev$. This behaviour can be explained by the strong interference between the $\rho(1450)^0$ and $\rho(1700)^0$ with the $K$\Kb channel~\cite{pelaez19}. 

The relatively large contribution of the two $\rho$-like resonances, approximately six times greater than that of the $\rho^0(770)$ contribution in the \Dsppp decay, is inconsistent with the interpretation of these states as the first radial and orbital excitations of the ground state $\rho^0(770)$ meson~\cite{PDG2022}. The $\rho(1450)^0$ and $\rho(1700)^0$ states are well established, but their nature is still uncertain.

The quark model predicts the existence of two spin-2 states with masses between 1.2 and 1.6\gev and the same quantum numbers, $J^{PC}=0^{++}$~\cite{PDG2022}. The mixing of the two SU(3) states gives rise to the physical mesons $f_2(1270)$ and $f'_2(1525)$. The latter has a dominant $s\bar s$ component and a small $d\bar d + u\bar u$ component in its wave function, which implies a small probability of the $f'_2(1525)$ meson to decay into a pair of pions. Conversely, the $f_2(1270)$ meson is mostly a $d\bar d + u\bar u$ state, with a small $s\bar s$ component in its wave function. This resonance should therefore decay mainly into a pair of pions. The assigned quark content of the two tensor states is consistent with the observed ratio of branching fractions (\%)~\cite{PDG2022},
\begin{equation*}
\frac{\mathcal{B}(f_2(1270)\to K\Kb)}{\mathcal{B} (f_2(1270)\to\pi\pi)} =\frac{4.6^{+0.5}_{-0.4}}{84.8^{+2.9}_{-0.9}},\hskip 1.cm \frac{\mathcal{B}(f'_2(1525)\to K\Kb)}{\mathcal{B} (f'_2(1525)\to\pi\pi)} =\frac{87.6\pm 2.2}{0.83\pm 0.16}~.
\end{equation*}

The very small fit fraction of the $f'_2(1525)\pi^+$ channel in the \Dsppp decay is consistent with this picture. One would expect the $f_2(1270)$ to be produced at a higher rate from the $d\bar d$ source in the \Dppp decay, compared to the rate from the $s\bar s$ source in the \Dsppp decay. Surprisingly, the fit fraction of the $f_2(1270)$ resonance is found to be the same in both decays.

\section{Final results and conclusions}
\label{sec:conclusion}

\begin{figure}[h] 
 \centering
\includegraphics[width= \textwidth]{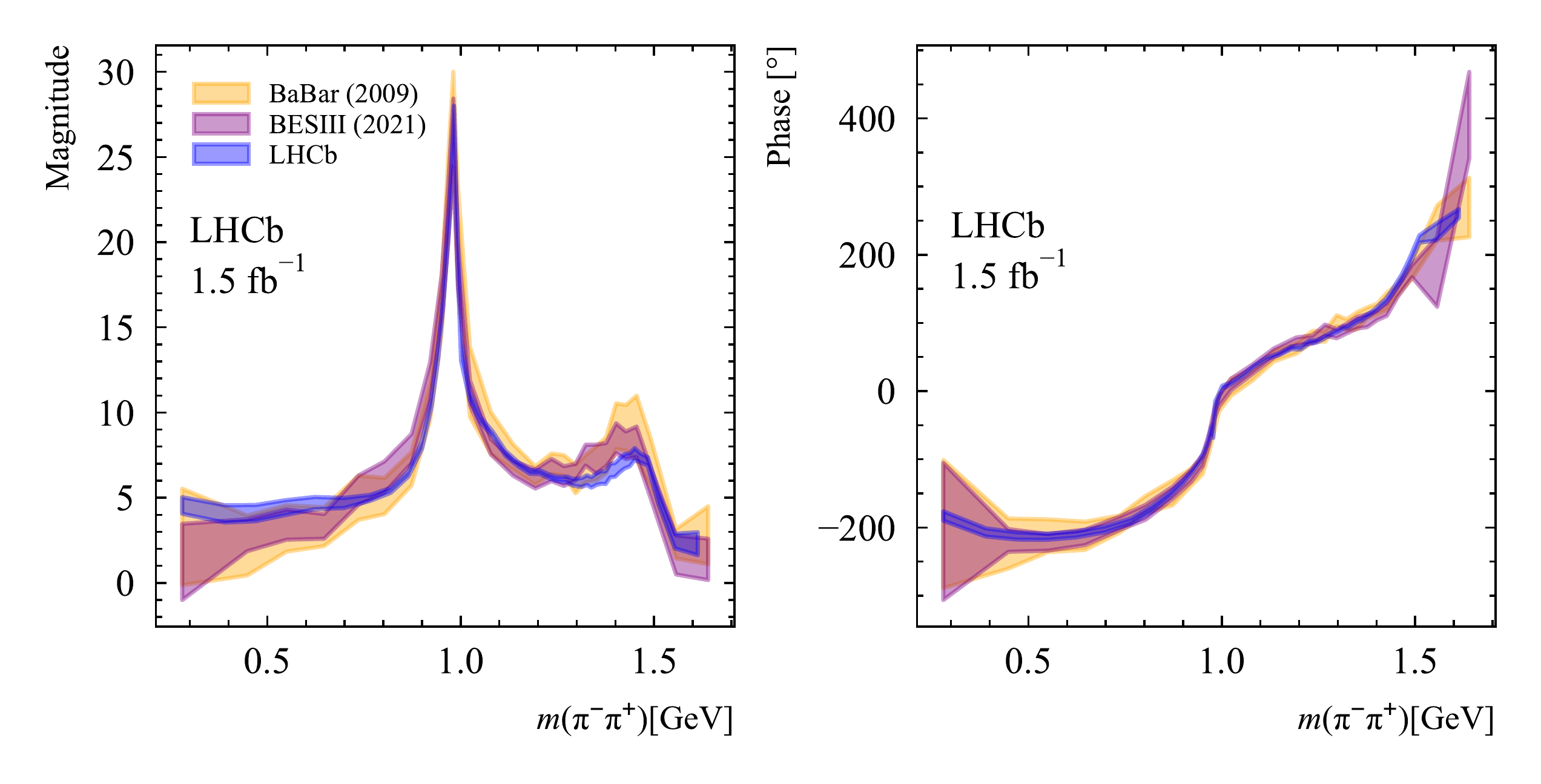}
\caption{Comparison of \pip\pim S-wave amplitude from the  \Dsppp decay with previous results from BaBar~\cite{babar-ds} and BESIII~\cite{besiiicollaboration2021amplitude}.}
\label{fig:DDsSW2}
\end{figure}

Based on a sample containing over $7\times10^5$ signal candidates with a purity of 95\%, a Dalitz plot analysis of the \Dsppp decay is performed and the resonant structure and the $\pi^+\pi^-$ amplitude in $S$-wave are determined. The data are described by a model with contributions from S-, P- and D-waves. 
The S-wave contribution accounts for nearly 85\% of the decay rate. The D-wave contribution contains two states, the $f_2(1270)$ and $f'_2(1525)$ resonances, and is the second largest component, with a fit fraction of 13.12\%. The P-wave contribution has four components, corresponding to the resonances $\rho(770)^0$, $\omega(782)$, $\rho(1450)^0$ and $\rho(1700)^0$, with a fit fraction of 8.55\%. The final results, including systematic uncertainties, are presented in Table~\ref{tab:fit-final}.  The results of this analysis are in agreement with previous measurements, as shown in Fig.~\ref{fig:DDsSW2} and Table~\ref{tab:ffs-final2}. The magnitudes and phases of the S-wave contribution are given in Table~\ref{tab:mi_pwa_1}.

\begin{table}[h!]
\centering
\caption{Final results of the \Dsppp Dalitz plot fit. The uncertainties are statistical, experimental systematic and associated to the decay amplitude model, respectively.}
\label{tab:fit-final}
\newcommand{\qc}[1]{\multicolumn{4}{c}{#1}}
\begin{tabular}{>{$}c<{$}| >{$}r<{\,\pm\,$}@{}>{$}r<{\,\pm\,$}@{}>{$}r<{\,\pm\,$}@{}>{$}l<{$} >{$}r<{\,\pm\,$}@{}>{$}r<{\,\pm\,$}@{}>{$}r<{\,\pm\,$}@{}>{$}l<{$}}
\toprule
\multicolumn{1}{c|}{Resonance} & \qc{Magnitude} & \qc{Phase $[^\circ]$} \\
\midrule
\rho(770)^{0}	 & 0.1201 &0.0030 &0.0050&0.0062  & 79.4  &1.8&7.8 &4.4 \\
\omega(782)	 & 0.04001&0.00090&0.0018&0.00086 & -109.9&1.7&0.94&1.4 \\
\rho(1450)^{0}   & 1.277  &0.026  &0.023 &0.48    & -115.2&2.6&2.8 &10  \\
\rho(1700)^{0}   & 0.873  &0.061  &0.054 &0.62    & -60.9 &6.1&6.7 &12  \\
f_2(1270)        & \qc{1 (fixed)}                 & \qc{0 (fixed)}      \\
f'_2(1525)      & 0.1098&0.0069  &0.019 &0.015   & 178.1 &4.2&12  &7   \\
\midrule[\heavyrulewidth]
& \qc{Fit Fraction (FF) [\%]} \\
\midrule
\multicolumn{1}{c|}{S-wave} & 84.97&0.14 &0.30 &0.63   \\
\rho(770)^{0}	              & 1.038&0.054&0.097&0.11   \\
\omega(782)	              & 0.360&0.016&0.034&0.016  \\
\midrule
\rho(1450)^{0}                & 3.86 &0.15 &0.14 &2.0    \\
\rho(1700)^{0}                & 0.365&0.050&0.045&0.34   \\
\multicolumn{1}{c|}{combined} & 6.14 &0.27 &0.34 &1.9    \\
\midrule
f_2(1270)                     & 13.69&0.14 &0.22 &0.49    \\
f'_2(1525)                   & 0.0528&0.0070&0.015&0.0087 \\ 
\bottomrule
\end{tabular}
\end{table}

\begin{table}[h!]
   \centering
    \caption{Resonant structure of the \Dsppp decay from this analysis compared to previous determinations from BaBar~\cite{babar-ds} and BESIII~\cite{besiiicollaboration2021amplitude}. The fit fractions are given in per cent. The statistical and systematic uncertainties are added in quadrature.\label{tab:ffs-final2}}
\newcommand{\dc}[1]{\multicolumn{2}{c}{#1}}
\begin{tabular}{>{$}c<{$}| >{$}r<{\,\pm\,$}@{}>{$}l<{$} >{$}r<{\,\pm\,$}@{}>{$}l<{$} >{$}r<{\,\pm\,$}@{}>{$}l<{$}}
\toprule
\multicolumn{1}{c|}{mode} &
\multicolumn{2}{c}{this result} &
\multicolumn{2}{c}{BaBar} &
\multicolumn{2}{c}{BESIII} \\
\midrule
\multicolumn{1}{c|}{S-wave} & 84.97&0.64  & 83.0&2.1 & 84.2&1.4 \\
\rho(770)^0\pi^+    &   1.04&0.12  & 1.8&1.1  & 0.9&0.8  \\
\omega(782)\pi^+    &  0.360&0.022 & \dc{--}  & \dc{--}  \\
\rho(1450)^{0}\pi^+ &   3.86&2.0   & 2.3&1.9  & 1.3&0.8  \\
\rho(1700)^{0}\pi^+ &   0.37&0.34  & \dc{--}  & \dc{--}  \\
f_2(1270)\pi^+      &  13.60&0.50  & 10.1&1.9 & 10.5&1.4 \\
f'_2(1525)\pi^+     &  0.045&0.011 & \dc{--}  & \dc{--}  \\
\bottomrule
\end{tabular}
\end{table}

\begin{table}
\caption{Magnitude and phase of the S-wave amplitude as a function of \pim\pip~invariant mass. The uncertainties are statistical, experimental and model, respectively. The $\pi^+\pi^-$ invariant mass is expressed in\gev.\label{tab:mi_pwa_1}}
\centering
\scalefont{0.8}
\begin{tabular}{ c| >{$}r<{\,\pm\,$}@{}>{$}r<{\,\pm\,$}@{}>{$}r<{\,\pm\,$}@{}>{$}l<{$} >{$}r<{\,\pm\,$}@{}>{$}r<{\,\pm\,$}@{}>{$}r<{\,\pm\,$}@{}>{$}l<{$}}
\toprule
 $m_{\pi\pi}$  & \multicolumn{4}{c}{Magnitude} & \multicolumn{4}{c}{Phase $[^\circ]$} \\
\midrule
   0.280   &   4.54  &0.15 &0.24 &0.46   &   176.8  &2.0 &3.5 &6.6    \\
   0.390   &   4.05  &0.12 &0.11 &0.50   &   152.8  &1.5 &2.3 &5.8    \\
   0.470   &   4.10  &0.12 &0.11 &0.46   &   147.6  &1.4 &2.1 &5.1    \\
   0.546   &   4.41  &0.11 &0.11 &0.42   &   146.4  &1.1 &1.7 &4.4    \\
   0.623   &   4.69  &0.10 &0.13 &0.34   &   149.6  &1.0 &1.3 &4.2    \\
   0.698   &   4.691 &0.092&0.14 &0.28   &   157.4  &1.0 &1.4 &4.3    \\
   0.766   &   4.994 &0.079&0.074&0.16   &   169.5  &1.1 &1.1 &3.8    \\
   0.819   &   5.43  &0.072&0.077&0.13   &   -172.8 &1.3 &1.3 &3.7    \\
   0.865   &   6.405 &0.066&0.068&0.14   &   -152.0 &1.2 &1.5 &2.7    \\
   0.900   &   8.096 &0.069&0.088&0.20   &   -133.0 &1.1 &2.3 &1.6    \\
   0.925   &   10.624&0.082&0.090&0.25   &   -116.5 &1.0 &1.8 &1.2    \\
   0.942   &   13.47 &0.10 &0.10 &0.29   &   -103.0 &1.0 &1.7 &0.8    \\
   0.955   &   16.56 &0.12 &0.13 &0.32   &   -88.8  &1.2 &1.6 &1.1    \\
   0.964   &   19.45 &0.14 &0.19 &0.42   &   -74.9  &1.2 &2.0 &0.95   \\
   0.972   &   22.15 &0.16 &0.27 &0.40   &   -59.5  &1.4 &1.5 &5.7    \\
   0.978   &   24.62 &0.17 &0.46 &0.68   &   -56.7  &1.4 &6.7 &11.8   \\
   0.983   &   26.95 &0.17 &0.42 &1.04   &   -21.8  &1.2 &4.2 &6.9    \\
   0.990   &   20.89 &0.14 &0.20 &0.90   &   -9.8   &1.1 &1.6 &2.8    \\
   1.001   &   13.695&0.091&0.13 &0.66   &   5.45   &0.92&1.5 &2.3    \\
   1.023   &   10.995&0.073&0.10 &0.22   &   11.28  &0.76&1.4 &1.3    \\
   1.051   &   9.593 &0.063&0.11 &0.13   &   21.57  &0.72&1.3 &0.93   \\
   1.083   &   8.731 &0.059&0.072&0.14   &   36.27  &0.65&1.4 &0.80   \\
   1.116   &   7.606 &0.059&0.038&0.13   &   48.02  &0.68&1.3 &0.66   \\
   1.149   &   6.961 &0.060&0.043&0.11   &   54.42  &0.71&0.93&1.2    \\
   1.179   &   6.515 &0.058&0.043&0.10   &   63.46  &0.79&0.82&2.0    \\
   1.205   &   6.506 &0.062&0.033&0.096  &   63.47  &0.79&0.66&2.05   \\
   1.227   &   6.264 &0.064&0.034&0.097  &   72.45  &0.89&0.78&2.0    \\
   1.245   &   6.125 &0.068&0.025&0.11   &   72.80  &0.94&0.68&1.4    \\
   1.261   &   6.081 &0.069&0.034&0.13   &   77.2   &1.0 &0.78&1.07   \\
   1.276   &   6.071 &0.072&0.036&0.15   &   82.2   &1.0 &0.43&0.95   \\
   1.289   &   5.912 &0.074&0.067&0.18   &   86.1   &1.0 &0.59&1.1    \\
   1.302   &   5.893 &0.078&0.099&0.22   &   88.8   &1.0 &0.48&1.1    \\
   1.314   &   5.901 &0.080&0.12 &0.25   &   93.2   &1.1 &0.55&1.3    \\
   1.326   &   6.031 &0.082&0.095&0.28   &   93.8   &1.1 &0.48&1.06   \\
   1.338   &   5.904 &0.083&0.067&0.30   &   98.7   &1.1 &0.61&1.1    \\
   1.351   &   6.086 &0.085&0.056&0.32   &   103.3  &1.0 &0.50&1.3    \\
   1.363   &   6.181 &0.089&0.075&0.35   &   105.7  &1.1 &0.65&1.5    \\
   1.375   &   6.185 &0.093&0.087&0.36   &   110.4  &1.1 &0.55&1.6    \\
   1.387   &   6.60  &0.10 &0.091&0.38   &   114.5  &1.0 &0.46&1.4    \\
   1.399   &   6.63  &0.10 &0.085&0.39   &   119.7  &1.0 &0.46&1.7    \\
   1.411   &   6.90  &0.11 &0.080&0.39   &   126.5  &1.0 &0.33&2.0    \\
   1.424   &   7.14  &0.11 &0.11 &0.41   &   132.3  &1.0 &0.37&2.4    \\
   1.437   &   7.22  &0.11 &0.11 &0.37   &   142.03 &0.92&0.47&3.0    \\
   1.450   &   7.56  &0.11 &0.12 &0.33   &   153.74 &0.86&0.60&3.5    \\
   1.465   &   7.33  &0.11 &0.13 &0.24   &   166.50 &0.83&0.68&4.1    \\
   1.484   &   7.13  &0.10 &0.15 &0.23   &   -172.15&0.82&0.81&4.9    \\
   1.511   &   5.009 &0.078&0.13 &0.37   &   -136.8 &1.1 &1.4 &5.2    \\
   1.554   &   2.456 &0.073&0.14 &0.43   &   -126.8 &2.7 &2.9 &12.0   \\
   1.613   &   2.31  &0.11 &0.14 &0.66   &   -99.5  &2.8 &5.9 &5.8    \\
   1.823   &   3.75  &0.27 &0.32 &0.63   &   3.0    &4.2 &6.3 &17     \\
\bottomrule
\end{tabular}
\end{table}

Significant differences are found between the resonant structure of the \Dsppp and \Dppp decays, as summarized in Table~\ref{tab:DsDp}. In the \Dp S-wave amplitude, a broad structure near threshold associated to the $f_0(500)$ resonance in model-dependent analyses~\cite{babar-ds,besiiicollaboration2021amplitude}, is observed. A corresponding structure is not found in the \Dsp decay. Conversely, the peak corresponding to the $f_0(980)$ meson is much more prominent in the \Dsp than in the \Dp decay. In both decays there is an indiciation of at least one scalar state at $m(\pi^+\pi^-)\sim$ $1.45$ GeV. Significant differences are also found between the phase of the S-wave and the phase shift $\delta_{0}^0$ of $\pi^+\pi^-\to\pi^+\pi^-$ scattering.

\begin{table}[h!]
\centering
\caption{Resonant structures of the \Dsppp and \Dppp~\cite{LHCb-PAPER-2022-016} decays, expressed fit fractions (\%). Uncertainties are only statistical.\label{tab:DsDp}}
\begin{tabular}{c >{$}r<{\,\pm\,$}@{}>{$}l<{$} >{$}r<{\,\pm\,$}@{}>{$}l<{$}}
\toprule
mode & \multicolumn{2}{c}{\Dsppp} & \multicolumn{2}{c}{\Dppp} \\
\midrule
S-wave & 84.97&0.14 & 61.82&0.5  \\
P-wave & 8.55 &0.44 & 32.31&0.64 \\
D-wave & 13.12&0.02 & 13.8 &0.2  \\
\bottomrule
\end{tabular}
\end{table}  

The observed differences in the S-wave could be explained by the hypothesis of the scalar resonances being produced by rescattering of pseudoscalar mesons. The latter could  originate from a $d\bar d$ source, in the case of the \Dp decay, and from an $s\bar s$ source in the case of the \Dsp decay. This mechanism would yield different sets of coupled channels, resulting in a different composition of the $\pi^+\pi^-$ S-wave component in \Dppp and \Dsppp decays. It would also explain the difference with respect to the $\delta_{0}^0$ phase shift of $\pi^+\pi^-\to\pi^+\pi^-$ scattering.

 The ratio between the $\rho(770)^0\pi^+$ and $\omega(782)\pi^+$ fit fractions is approximately 3 in the \Dsppp decay, compared to 250 in \Dppp decay. These results indicate that  the $\omega(782)$ is produced by different mechanisms in \Dsp and \Dp decays. In contrast to the large difference in the $\rho(770)^0\pi^+$ fit fractions, the combined contributions of the $\rho(1450)^0\pi^+$ and $\rho(1700)^0\pi^+$ channels are very similar, ($6.14\pm 0.27$)\% in \Dsppp decay and ($7.1\pm 0.8$)\% in \Dppp decay. These results challenge the interpretation of the $\rho(1450)^0$ and $\rho(1700)^0$ as excitations of the ground state $\rho(770)^0$. The same fit fraction of the $f_2(1270)\pi^+$ mode is measured in both decays. This is a surprising result, since one would expect the $f_2(1270)$ to be produced at a higher rate from a $d\bar d$ than from an $s\bar s$ source.

The determination of the $\pi^+\pi^-$ S-wave from \Dsppp decay provides an important input to phenomenological analyses, from which the scattering amplitudes could be obtained. The comparison between the resonant structure of the $D^+_s\to \pi^-\pi^+\pi^+$ and $D^+\to \pi^-\pi^+\pi^+$ decays provides valuable information for the understanding of the hadron formation mechanisms in charm meson decays.

\section*{Acknowledgements}
%
%
\noindent We express our gratitude to our colleagues in the CERN
accelerator departments for the excellent performance of the LHC. We
thank the technical and administrative staff at the LHCb
institutes.
We acknowledge support from CERN and from the national agencies:
CAPES, CNPq, FAPERJ and FINEP (Brazil); 
MOST and NSFC (China); 
CNRS/IN2P3 (France); 
BMBF, DFG and MPG (Germany); 
INFN (Italy); 
NWO (Netherlands); 
MNiSW and NCN (Poland); 
MEN/IFA (Romania); 
MICINN (Spain); 
SNSF and SER (Switzerland); 
NASU (Ukraine); 
STFC (United Kingdom); 
DOE NP and NSF (USA).
We acknowledge the computing resources that are provided by CERN, IN2P3
(France), KIT and DESY (Germany), INFN (Italy), SURF (Netherlands),
PIC (Spain), GridPP (United Kingdom), 
CSCS (Switzerland), IFIN-HH (Romania), CBPF (Brazil),
Polish WLCG  (Poland) and NERSC (USA).
We are indebted to the communities behind the multiple open-source
software packages on which we depend.
Individual groups or members have received support from
ARC and ARDC (Australia);
Minciencias (Colombia);
AvH Foundation (Germany);
EPLANET, Marie Sk\l{}odowska-Curie Actions and ERC (European Union);
A*MIDEX, ANR, IPhU and Labex P2IO, and R\'{e}gion Auvergne-Rh\^{o}ne-Alpes (France);
Key Research Program of Frontier Sciences of CAS, CAS PIFI, CAS CCEPP, 
Fundamental Research Funds for the Central Universities, 
and Sci. \& Tech. Program of Guangzhou (China);
GVA, XuntaGal, GENCAT and Prog.~Atracci\'on Talento, CM (Spain);
SRC (Sweden);
the Leverhulme Trust, the Royal Society
 and UKRI (United Kingdom).




\addcontentsline{toc}{section}{References}
\bibliographystyle{LHCb}
\bibliography{main,standard,LHCb-PAPER,LHCb-CONF,LHCb-DP,LHCb-TDR}

\ifx\mcitethebibliography\mciteundefinedmacro
\PackageError{LHCb.bst}{mciteplus.sty has not been loaded}
{This bibstyle requires the use of the mciteplus package.}\fi
\providecommand{\href}[2]{#2}
\begin{mcitethebibliography}{10}
\mciteSetBstSublistMode{n}
\mciteSetBstMaxWidthForm{subitem}{\alph{mcitesubitemcount})}
\mciteSetBstSublistLabelBeginEnd{\mcitemaxwidthsubitemform\space}
{\relax}{\relax}

\bibitem{asner}
D.~Asner, \ifthenelse{\boolean{articletitles}}{\emph{Charm dalitz plot analysis
  formalism and results},
  }{}\href{http://arxiv.org/abs/hep-ex/0410014}{{\normalfont\ttfamily
  arXiv:hep-ex/0410014}}\relax
\mciteBstWouldAddEndPuncttrue
\mciteSetBstMidEndSepPunct{\mcitedefaultmidpunct}
{\mcitedefaultendpunct}{\mcitedefaultseppunct}\relax
\EndOfBibitem
\bibitem{PDG2022}
Particle Data Group, R.~L. Workman {\em et~al.},
  \ifthenelse{\boolean{articletitles}}{\emph{{\href{http://pdg.lbl.gov/}{Review
  of particle physics}}}, }{}\href{https://doi.org/10.1093/ptep/ptac097}{Prog.\
  Theor.\ Exp.\ Phys.\  \textbf{2022} (2022) 083C01}\relax
\mciteBstWouldAddEndPuncttrue
\mciteSetBstMidEndSepPunct{\mcitedefaultmidpunct}
{\mcitedefaultendpunct}{\mcitedefaultseppunct}\relax
\EndOfBibitem
\bibitem{TripleM}
R.~T. Aoude, P.~C. Magalh\~aes, A.~C. dos Reis, and M.~R. Robilotta,
  \ifthenelse{\boolean{articletitles}}{\emph{Multimeson model for the
  ${D}^{+}\ensuremath{\rightarrow}{K}^{+}{K}^{\ensuremath{-}}{K}^{+}$ decay
  amplitude}, }{}\href{https://doi.org/10.1103/PhysRevD.98.056021}{Phys.\ Rev.\
   \textbf{D98} (2018) 056021}\relax
\mciteBstWouldAddEndPuncttrue
\mciteSetBstMidEndSepPunct{\mcitedefaultmidpunct}
{\mcitedefaultendpunct}{\mcitedefaultseppunct}\relax
\EndOfBibitem
\bibitem{LHCb-PAPER-2022-016}
LHCb collaboration, R.~Aaij {\em et~al.},
  \ifthenelse{\boolean{articletitles}}{\emph{{Amplitude analysis of the $D^+
  \to \pi^-\pi^+\pi^+$ decay and measurement of the $\pi^-\pi^+$ S-wave
  amplitude}}, }{}\href{http://arxiv.org/abs/2208.03300}{{\normalfont\ttfamily
  arXiv:2208.03300}}, {Submitted to JHEP}\relax
\mciteBstWouldAddEndPuncttrue
\mciteSetBstMidEndSepPunct{\mcitedefaultmidpunct}
{\mcitedefaultendpunct}{\mcitedefaultseppunct}\relax
\EndOfBibitem
\bibitem{Ochs_2013}
W.~Ochs, \ifthenelse{\boolean{articletitles}}{\emph{The status of glueballs},
  }{}\href{https://doi.org/10.1088/0954-3899/40/4/043001}{Journal of Physics G:
  Nuclear and Particle Physics \textbf{40} (2013) 043001}\relax
\mciteBstWouldAddEndPuncttrue
\mciteSetBstMidEndSepPunct{\mcitedefaultmidpunct}
{\mcitedefaultendpunct}{\mcitedefaultseppunct}\relax
\EndOfBibitem
\bibitem{babar-ds}
BaBar collaboration, B.~Aubert {\em et~al.},
  \ifthenelse{\boolean{articletitles}}{\emph{{Dalitz plot analysis of $D_s^+\to
  \pi^-\pi^+\pi^+$}},
  }{}\href{https://doi.org/http://dx.doi.org/10.1103/PhysRevD.79.032003}{Phys.\
  Rev.\  \textbf{D79} (2009) 032003}\relax
\mciteBstWouldAddEndPuncttrue
\mciteSetBstMidEndSepPunct{\mcitedefaultmidpunct}
{\mcitedefaultendpunct}{\mcitedefaultseppunct}\relax
\EndOfBibitem
\bibitem{besiiicollaboration2021amplitude}
BESIII collaboration, M.~Ablikim,  {\em et~al.},
  \ifthenelse{\boolean{articletitles}}{\emph{{Amplitude analysis of the
  $D_s^{+} \rightarrow \pi^{+} \pi^{-} \pi^{+}$ decay}},
  }{}\href{http://arxiv.org/abs/2108.10050}{{\normalfont\ttfamily
  arXiv:2108.10050}}\relax
\mciteBstWouldAddEndPuncttrue
\mciteSetBstMidEndSepPunct{\mcitedefaultmidpunct}
{\mcitedefaultendpunct}{\mcitedefaultseppunct}\relax
\EndOfBibitem
\bibitem{e791Dp3pi}
E791 collaboration, E.~M. Aitala {\em et~al.},
  \ifthenelse{\boolean{articletitles}}{\emph{{Experimental evidence for a light
  and broad scalar resonance in $D^+\to \pi^-\pi^+\pi^+$ decay}},
  }{}\href{https://doi.org/10.1103/PhysRevLett.86.770}{Phys.\ Rev.\ Lett
  \textbf{86} (2001) 770}\relax
\mciteBstWouldAddEndPuncttrue
\mciteSetBstMidEndSepPunct{\mcitedefaultmidpunct}
{\mcitedefaultendpunct}{\mcitedefaultseppunct}\relax
\EndOfBibitem
\bibitem{focus3pi}
FOCUS collaboration, J.~M. Link {\em et~al.},
  \ifthenelse{\boolean{articletitles}}{\emph{{Dalitz plot analysis of $D_s^+$
  and $D^+$ decay to $\pi^+ \pi^- \pi^+$ using the K-matrix formalism}},
  }{}\href{https://doi.org/10.1016/j.physletb.2004.01.065}{Phys.\ Lett.\
  \textbf{B585} (2004) 200}\relax
\mciteBstWouldAddEndPuncttrue
\mciteSetBstMidEndSepPunct{\mcitedefaultmidpunct}
{\mcitedefaultendpunct}{\mcitedefaultseppunct}\relax
\EndOfBibitem
\bibitem{CleoD3pi}
CLEO collaboration, G.~Bonvicini {\em et~al.},
  \ifthenelse{\boolean{articletitles}}{\emph{{Dalitz plot analisis of the \Dppp
  decay}}, }{}\href{https://doi.org/10.1103/PhysRevD.76.012001}{Phys.\ Rev.\
  \textbf{D76} (2007) 012001}\relax
\mciteBstWouldAddEndPuncttrue
\mciteSetBstMidEndSepPunct{\mcitedefaultmidpunct}
{\mcitedefaultendpunct}{\mcitedefaultseppunct}\relax
\EndOfBibitem
\bibitem{LHCb_detector2008}
LHCb collaboration, A.~A. Alves~Jr.\ {\em et~al.},
  \ifthenelse{\boolean{articletitles}}{\emph{{The LHCb Detector at the LHC}},
  }{}\href{https://doi.org/10.1088/1748-0221/3/08/S08005}{JINST \textbf{3}
  (2008) S08005}\relax
\mciteBstWouldAddEndPuncttrue
\mciteSetBstMidEndSepPunct{\mcitedefaultmidpunct}
{\mcitedefaultendpunct}{\mcitedefaultseppunct}\relax
\EndOfBibitem
\bibitem{LHCb-DP-2014-002}
LHCb collaboration, R.~Aaij {\em et~al.},
  \ifthenelse{\boolean{articletitles}}{\emph{{LHCb detector performance}},
  }{}\href{https://doi.org/10.1142/S0217751X15300227}{Int.\ J.\ Mod.\ Phys.\
  \textbf{A30} (2015) 1530022},
  \href{http://arxiv.org/abs/1412.6352}{{\normalfont\ttfamily
  arXiv:1412.6352}}\relax
\mciteBstWouldAddEndPuncttrue
\mciteSetBstMidEndSepPunct{\mcitedefaultmidpunct}
{\mcitedefaultendpunct}{\mcitedefaultseppunct}\relax
\EndOfBibitem
\bibitem{Sjostrand:2007gs}
T.~Sj\"{o}strand, S.~Mrenna, and P.~Skands,
  \ifthenelse{\boolean{articletitles}}{\emph{{A brief introduction to PYTHIA
  8.1}}, }{}\href{https://doi.org/10.1016/j.cpc.2008.01.036}{Comput.\ Phys.\
  Commun.\  \textbf{178} (2008) 852},
  \href{http://arxiv.org/abs/0710.3820}{{\normalfont\ttfamily
  arXiv:0710.3820}}\relax
\mciteBstWouldAddEndPuncttrue
\mciteSetBstMidEndSepPunct{\mcitedefaultmidpunct}
{\mcitedefaultendpunct}{\mcitedefaultseppunct}\relax
\EndOfBibitem
\bibitem{Sjostrand:2006za}
T.~Sj\"{o}strand, S.~Mrenna, and P.~Skands,
  \ifthenelse{\boolean{articletitles}}{\emph{{PYTHIA 6.4 physics and manual}},
  }{}\href{https://doi.org/10.1088/1126-6708/2006/05/026}{JHEP \textbf{05}
  (2006) 026}, \href{http://arxiv.org/abs/hep-ph/0603175}{{\normalfont\ttfamily
  arXiv:hep-ph/0603175}}\relax
\mciteBstWouldAddEndPuncttrue
\mciteSetBstMidEndSepPunct{\mcitedefaultmidpunct}
{\mcitedefaultendpunct}{\mcitedefaultseppunct}\relax
\EndOfBibitem
\bibitem{LHCb-PROC-2010-056}
I.~Belyaev {\em et~al.}, \ifthenelse{\boolean{articletitles}}{\emph{{Handling
  of the generation of primary events in Gauss, the LHCb simulation
  framework}}, }{}\href{https://doi.org/10.1088/1742-6596/331/3/032047}{J.\
  Phys.\ Conf.\ Ser.\  \textbf{331} (2011) 032047}\relax
\mciteBstWouldAddEndPuncttrue
\mciteSetBstMidEndSepPunct{\mcitedefaultmidpunct}
{\mcitedefaultendpunct}{\mcitedefaultseppunct}\relax
\EndOfBibitem
\bibitem{Lange:2001uf}
D.~J. Lange, \ifthenelse{\boolean{articletitles}}{\emph{{The EvtGen particle
  decay simulation package}},
  }{}\href{https://doi.org/10.1016/S0168-9002(01)00089-4}{Nucl.\ Instrum.\
  Meth.\  \textbf{A462} (2001) 152}\relax
\mciteBstWouldAddEndPuncttrue
\mciteSetBstMidEndSepPunct{\mcitedefaultmidpunct}
{\mcitedefaultendpunct}{\mcitedefaultseppunct}\relax
\EndOfBibitem
\bibitem{davidson2015photos}
N.~Davidson, T.~Przedzinski, and Z.~Was,
  \ifthenelse{\boolean{articletitles}}{\emph{{PHOTOS interface in C++:
  Technical and physics documentation}},
  }{}\href{https://doi.org/https://doi.org/10.1016/j.cpc.2015.09.013}{Comp.\
  Phys.\ Comm.\  \textbf{199} (2016) 86},
  \href{http://arxiv.org/abs/1011.0937}{{\normalfont\ttfamily
  arXiv:1011.0937}}\relax
\mciteBstWouldAddEndPuncttrue
\mciteSetBstMidEndSepPunct{\mcitedefaultmidpunct}
{\mcitedefaultendpunct}{\mcitedefaultseppunct}\relax
\EndOfBibitem
\bibitem{Allison:2006ve}
Geant4 collaboration, J.~Allison {\em et~al.},
  \ifthenelse{\boolean{articletitles}}{\emph{{Geant4 developments and
  applications}}, }{}\href{https://doi.org/10.1109/TNS.2006.869826}{IEEE
  Trans.\ Nucl.\ Sci.\  \textbf{53} (2006) 270}\relax
\mciteBstWouldAddEndPuncttrue
\mciteSetBstMidEndSepPunct{\mcitedefaultmidpunct}
{\mcitedefaultendpunct}{\mcitedefaultseppunct}\relax
\EndOfBibitem
\bibitem{Agostinelli:2002hh}
Geant4 collaboration, S.~Agostinelli {\em et~al.},
  \ifthenelse{\boolean{articletitles}}{\emph{{Geant4: A simulation toolkit}},
  }{}\href{https://doi.org/10.1016/S0168-9002(03)01368-8}{Nucl.\ Instrum.\
  Meth.\  \textbf{A506} (2003) 250}\relax
\mciteBstWouldAddEndPuncttrue
\mciteSetBstMidEndSepPunct{\mcitedefaultmidpunct}
{\mcitedefaultendpunct}{\mcitedefaultseppunct}\relax
\EndOfBibitem
\bibitem{LHCb-PROC-2011-006}
M.~Clemencic {\em et~al.}, \ifthenelse{\boolean{articletitles}}{\emph{{The
  \lhcb simulation application, Gauss: Design, evolution and experience}},
  }{}\href{https://doi.org/10.1088/1742-6596/331/3/032023}{J.\ Phys.\ Conf.\
  Ser.\  \textbf{331} (2011) 032023}\relax
\mciteBstWouldAddEndPuncttrue
\mciteSetBstMidEndSepPunct{\mcitedefaultmidpunct}
{\mcitedefaultendpunct}{\mcitedefaultseppunct}\relax
\EndOfBibitem
\bibitem{Breiman}
L.~Breiman, J.~H. Friedman, R.~A. Olshen, and C.~J. Stone, {\em Classification
  and regression trees}, Wadsworth international group, Belmont, California,
  USA, 1984\relax
\mciteBstWouldAddEndPuncttrue
\mciteSetBstMidEndSepPunct{\mcitedefaultmidpunct}
{\mcitedefaultendpunct}{\mcitedefaultseppunct}\relax
\EndOfBibitem
\bibitem{AdaBoost}
Y.~Freund and R.~E. Schapire, \ifthenelse{\boolean{articletitles}}{\emph{A
  decision-theoretic generalization of on-line learning and an application to
  boosting}, }{}\href{https://doi.org/10.1006/jcss.1997.1504}{J.\ Comput.\
  Syst.\ Sci.\  \textbf{55} (1997) 119}\relax
\mciteBstWouldAddEndPuncttrue
\mciteSetBstMidEndSepPunct{\mcitedefaultmidpunct}
{\mcitedefaultendpunct}{\mcitedefaultseppunct}\relax
\EndOfBibitem
\bibitem{rogozhnikov2016reweighting}
A.~Rogozhnikov, \ifthenelse{\boolean{articletitles}}{\emph{Reweighting with
  boosted decision trees}, }{} in {\em Journal of Physics: Conference Series},
  \textbf{762}, 012036, IOP Publishing, 2016\relax
\mciteBstWouldAddEndPuncttrue
\mciteSetBstMidEndSepPunct{\mcitedefaultmidpunct}
{\mcitedefaultendpunct}{\mcitedefaultseppunct}\relax
\EndOfBibitem
\bibitem{Pivk:2004ty}
M.~Pivk and F.~R. Le~Diberder,
  \ifthenelse{\boolean{articletitles}}{\emph{{sPlot: A statistical tool to
  unfold data distributions}},
  }{}\href{https://doi.org/10.1016/j.nima.2005.08.106}{Nucl.\ Instrum.\ Meth.\
  \textbf{A555} (2005) 356},
  \href{http://arxiv.org/abs/physics/0402083}{{\normalfont\ttfamily
  arXiv:physics/0402083}}\relax
\mciteBstWouldAddEndPuncttrue
\mciteSetBstMidEndSepPunct{\mcitedefaultmidpunct}
{\mcitedefaultendpunct}{\mcitedefaultseppunct}\relax
\EndOfBibitem
\bibitem{das2016simple}
S.~Das, \ifthenelse{\boolean{articletitles}}{\emph{{A simple alternative to the
  Crystal Ball function}},
  }{}\href{http://arxiv.org/abs/1603.08591}{{\normalfont\ttfamily
  arXiv:1603.08591}}\relax
\mciteBstWouldAddEndPuncttrue
\mciteSetBstMidEndSepPunct{\mcitedefaultmidpunct}
{\mcitedefaultendpunct}{\mcitedefaultseppunct}\relax
\EndOfBibitem
\bibitem{book:BlattWeisskopf}
J.~Blatt and J.~Weisskopf, {\em Theoretical Nuclear Physics}, John Wiley and
  Sons, 1952\relax
\mciteBstWouldAddEndPuncttrue
\mciteSetBstMidEndSepPunct{\mcitedefaultmidpunct}
{\mcitedefaultendpunct}{\mcitedefaultseppunct}\relax
\EndOfBibitem
\bibitem{gounaris1968finite}
G.~Gounaris and J.~Sakurai,
  \ifthenelse{\boolean{articletitles}}{\emph{Finite-width corrections to the
  vector-meson-dominance prediction for $\rho \rightarrow e^+ e^-$}, }{}Phys.\
  Rev.\ Lett.\  \textbf{21} (1968) 244\relax
\mciteBstWouldAddEndPuncttrue
\mciteSetBstMidEndSepPunct{\mcitedefaultmidpunct}
{\mcitedefaultendpunct}{\mcitedefaultseppunct}\relax
\EndOfBibitem
\bibitem{ABRAROV20111894}
S.~M. Abrarov and B.~M. Quine,
  \ifthenelse{\boolean{articletitles}}{\emph{Efficient algorithmic
  implementation of the voigt/complex error function based on exponential
  series approximation},
  }{}\href{https://doi.org/https://doi.org/10.1016/j.amc.2011.06.072}{Applied
  Mathematics and Computation \textbf{218} (2011) 1894}\relax
\mciteBstWouldAddEndPuncttrue
\mciteSetBstMidEndSepPunct{\mcitedefaultmidpunct}
{\mcitedefaultendpunct}{\mcitedefaultseppunct}\relax
\EndOfBibitem
\bibitem{LHCb-PAPER-2014-012}
LHCb collaboration, R.~Aaij {\em et~al.},
  \ifthenelse{\boolean{articletitles}}{\emph{{Measurement of the resonant and
  \CP components in \mbox{\decay{\Bzb}{\jpsi\pip\pim}} decays}},
  }{}\href{https://doi.org/10.1103/PhysRevD.90.012003}{Phys.\ Rev.\
  \textbf{D90} (2014) 012003},
  \href{http://arxiv.org/abs/1404.5673}{{\normalfont\ttfamily
  arXiv:1404.5673}}\relax
\mciteBstWouldAddEndPuncttrue
\mciteSetBstMidEndSepPunct{\mcitedefaultmidpunct}
{\mcitedefaultendpunct}{\mcitedefaultseppunct}\relax
\EndOfBibitem
\bibitem{LHCB-PAPER-2013-069}
LHCb collaboration, R.~Aaij {\em et~al.},
  \ifthenelse{\boolean{articletitles}}{\emph{{Measurement of resonant and \CP
  components in \mbox{\decay{\Bsb}{\jpsi\pip\pim}} decays}},
  }{}\href{https://doi.org/10.1103/PhysRevD.89.092006}{Phys.\ Rev.\
  \textbf{D89} (2014) 092006},
  \href{http://arxiv.org/abs/1402.6248}{{\normalfont\ttfamily
  arXiv:1402.6248}}\relax
\mciteBstWouldAddEndPuncttrue
\mciteSetBstMidEndSepPunct{\mcitedefaultmidpunct}
{\mcitedefaultendpunct}{\mcitedefaultseppunct}\relax
\EndOfBibitem
\bibitem{PDG2020}
Particle Data Group, P.~A. Zyla {\em et~al.},
  \ifthenelse{\boolean{articletitles}}{\emph{{\href{http://pdg.lbl.gov/}{Review
  of particle physics}}}, }{}\href{https://doi.org/10.1093/ptep/ptaa104}{Prog.\
  Theor.\ Exp.\ Phys.\  \textbf{2020} (2020) 083C01}\relax
\mciteBstWouldAddEndPuncttrue
\mciteSetBstMidEndSepPunct{\mcitedefaultmidpunct}
{\mcitedefaultendpunct}{\mcitedefaultseppunct}\relax
\EndOfBibitem
\bibitem{paper:kaminski2008pion}
R.~Kami{\'n}ski, J.~Pelaez, and F.~Yndurain,
  \ifthenelse{\boolean{articletitles}}{\emph{Pion-pion scattering amplitude.
  iii. improving the analysis with forward dispersion relations and roy
  equations}, }{}Phys.\ Rev.\  \textbf{D77} (2008) 054015\relax
\mciteBstWouldAddEndPuncttrue
\mciteSetBstMidEndSepPunct{\mcitedefaultmidpunct}
{\mcitedefaultendpunct}{\mcitedefaultseppunct}\relax
\EndOfBibitem
\bibitem{paper:oset}
E.~Oset {\em et~al.}, \ifthenelse{\boolean{articletitles}}{\emph{Weak decays of
  heavy hadrons into dynamically generated resonances},
  }{}\href{https://doi.org/https://doi.org/10.1142/S0218301316300010}{Int.\ J.\
  Mod.\ Phys.\  \textbf{25} (2016) 1630001}\relax
\mciteBstWouldAddEndPuncttrue
\mciteSetBstMidEndSepPunct{\mcitedefaultmidpunct}
{\mcitedefaultendpunct}{\mcitedefaultseppunct}\relax
\EndOfBibitem
\bibitem{hyams}
B.~Hyams {\em et~al.}, \ifthenelse{\boolean{articletitles}}{\emph{$\pi\pi$
  phase-shift analysis from 600 to 1900 mev}, }{}Nucl.\ Phys.\  \textbf{B64}
  (1973) 134\relax
\mciteBstWouldAddEndPuncttrue
\mciteSetBstMidEndSepPunct{\mcitedefaultmidpunct}
{\mcitedefaultendpunct}{\mcitedefaultseppunct}\relax
\EndOfBibitem
\bibitem{NA48}
NA48/2 collaboration, J.~R. Batley {\em et~al.},
  \ifthenelse{\boolean{articletitles}}{\emph{{Precise tests of low-energy QCD
  from $K_{e4}$ decay properties}},
  }{}\href{https://doi.org/10.1140/epjc/s10052-010-1480-6}{Eur.\ Phys.\ J.\
  \textbf{C70} (2010) 635}\relax
\mciteBstWouldAddEndPuncttrue
\mciteSetBstMidEndSepPunct{\mcitedefaultmidpunct}
{\mcitedefaultendpunct}{\mcitedefaultseppunct}\relax
\EndOfBibitem
\bibitem{watson1952effect}
K.~M. Watson, \ifthenelse{\boolean{articletitles}}{\emph{The effect of final
  state interactions on reaction cross sections}, }{}Phys.\ Rev.\  \textbf{88}
  (1952) 1163\relax
\mciteBstWouldAddEndPuncttrue
\mciteSetBstMidEndSepPunct{\mcitedefaultmidpunct}
{\mcitedefaultendpunct}{\mcitedefaultseppunct}\relax
\EndOfBibitem
\bibitem{gourdin}
M.~Gourdin {\em et~al.},
  \ifthenelse{\boolean{articletitles}}{\emph{Electromagnetic mixing of $\rho$,
  $\omega$ mesons},
  }{}\href{https://doi.org/https://doi.org/10.1016/0370-2693(69)90502-4}{Phys.\
  Lett.\  \textbf{B30} (1969) 347}\relax
\mciteBstWouldAddEndPuncttrue
\mciteSetBstMidEndSepPunct{\mcitedefaultmidpunct}
{\mcitedefaultendpunct}{\mcitedefaultseppunct}\relax
\EndOfBibitem
\bibitem{CBarrel-rw}
Crystal Barrel collaboration, A.~Abele {\em et~al.},
  \ifthenelse{\boolean{articletitles}}{\emph{{$\rho-\omega$ interference $\bar
  p p$ annihilation at rest into $\pi^-\pi^-\eta$}}, }{}Physics Letters
  \textbf{B411} (1997) 354\relax
\mciteBstWouldAddEndPuncttrue
\mciteSetBstMidEndSepPunct{\mcitedefaultmidpunct}
{\mcitedefaultendpunct}{\mcitedefaultseppunct}\relax
\EndOfBibitem
\bibitem{CMD}
R.~Akhmetshina {\em et~al.},
  \ifthenelse{\boolean{articletitles}}{\emph{Measurement of $e^+ e^-
  \rightarrow \pi^+\pi^-$ cross section with {CMD}-2 around $\rho$-meson},
  }{}\href{https://doi.org/https://doi.org/10.1016/S0370-2693(02)01168-1}{Phys.\
  Lett.\  \textbf{B527} (2002) 161},
  \href{http://arxiv.org/abs/0112031}{{\normalfont\ttfamily
  arXiv:0112031}}\relax
\mciteBstWouldAddEndPuncttrue
\mciteSetBstMidEndSepPunct{\mcitedefaultmidpunct}
{\mcitedefaultendpunct}{\mcitedefaultseppunct}\relax
\EndOfBibitem
\bibitem{LHCb-PAPER-2019-017}
LHCb collaboration, R.~Aaij {\em et~al.},
  \ifthenelse{\boolean{articletitles}}{\emph{{Amplitude analysis of the
  \mbox{\decay{\Bp}{\pip\pip\pim}} decay}},
  }{}\href{https://doi.org/10.1103/PhysRevD.101.012006}{Phys.\ Rev.\
  \textbf{D101} (2020) 012006},
  \href{http://arxiv.org/abs/1909.05211}{{\normalfont\ttfamily
  arXiv:1909.05211}}\relax
\mciteBstWouldAddEndPuncttrue
\mciteSetBstMidEndSepPunct{\mcitedefaultmidpunct}
{\mcitedefaultendpunct}{\mcitedefaultseppunct}\relax
\EndOfBibitem
\bibitem{LHCb-PAPER-2021-045}
LHCb collaboration, R.~Aaij {\em et~al.},
  \ifthenelse{\boolean{articletitles}}{\emph{{Observation of sizeable $\omega$
  contribution to \mbox{$\chicone \to \pi^+\pi^- J/\psi$ decays}}},
  }{}\href{http://arxiv.org/abs/2204.12597}{{\normalfont\ttfamily
  arXiv:2204.12597}}, {submitted to PRL}\relax
\mciteBstWouldAddEndPuncttrue
\mciteSetBstMidEndSepPunct{\mcitedefaultmidpunct}
{\mcitedefaultendpunct}{\mcitedefaultseppunct}\relax
\EndOfBibitem
\bibitem{KLOE:2003kas}
KLOE collaboration, A.~Aloisio {\em et~al.},
  \ifthenelse{\boolean{articletitles}}{\emph{{Study of the decay $\phi \to
  \pi^+ \pi^- \pi^0$ with the KLOE detector}},
  }{}\href{https://doi.org/10.1016/j.physletb.2005.01.092}{Phys.\ Lett.\
  \textbf{B561} (2003) 55},
  \href{http://arxiv.org/abs/hep-ex/0303016}{{\normalfont\ttfamily
  arXiv:hep-ex/0303016}}, [Erratum: Phys. Lett. \textbf{B609}, 449--450
  (2005)]\relax
\mciteBstWouldAddEndPuncttrue
\mciteSetBstMidEndSepPunct{\mcitedefaultmidpunct}
{\mcitedefaultendpunct}{\mcitedefaultseppunct}\relax
\EndOfBibitem
\bibitem{cheng}
H.~Cheng and C.~Chiang, \ifthenelse{\boolean{articletitles}}{\emph{{Two-body
  hadronic charmed meson decays}},
  }{}\href{https://doi.org/https://doi.org/10.1103/PhysRevD.81.074021}{Phys.\
  Rev.\  \textbf{D81} (2010) 074021}\relax
\mciteBstWouldAddEndPuncttrue
\mciteSetBstMidEndSepPunct{\mcitedefaultmidpunct}
{\mcitedefaultendpunct}{\mcitedefaultseppunct}\relax
\EndOfBibitem
\bibitem{yu}
Y.~Yu, Y.~Hsiao, and B.~Ke, \ifthenelse{\boolean{articletitles}}{\emph{{Study
  of the $D^+_s\to a_0(980)\rho$ and $a_0(980)\omega$} decays},
  }{}\href{http://arxiv.org/abs/2108.02936}{{\normalfont\ttfamily
  arXiv:2108.02936}}\relax
\mciteBstWouldAddEndPuncttrue
\mciteSetBstMidEndSepPunct{\mcitedefaultmidpunct}
{\mcitedefaultendpunct}{\mcitedefaultseppunct}\relax
\EndOfBibitem
\bibitem{cbarrel1}
Crystal Barrel collaboration, A.~Abele {\em et~al.},
  \ifthenelse{\boolean{articletitles}}{\emph{{High mass $\rho$-meson states
  from $\bar p d$-annihilation at rest into $\pi^-\pi^0\pi^0 p_{\mathrm{
  spectator}}$}}, }{}Phys.\ Lett.\  \textbf{B391} (1997) 191\relax
\mciteBstWouldAddEndPuncttrue
\mciteSetBstMidEndSepPunct{\mcitedefaultmidpunct}
{\mcitedefaultendpunct}{\mcitedefaultseppunct}\relax
\EndOfBibitem
\bibitem{obelix}
Obelix collaboration, A.~Bertin {\em et~al.},
  \ifthenelse{\boolean{articletitles}}{\emph{{Study of isovector scalar mesons
  $\rho$-meson states in the channel $\bar p \to K^{\pm}K^0_S\pi^{\mp} $} at
  rest with initial angular momentum state selection}, }{}Phys.\ Lett.\
  \textbf{B434} (1998) 180\relax
\mciteBstWouldAddEndPuncttrue
\mciteSetBstMidEndSepPunct{\mcitedefaultmidpunct}
{\mcitedefaultendpunct}{\mcitedefaultseppunct}\relax
\EndOfBibitem
\bibitem{paper:belle-tau}
Belle collaboration, M.~Fujikawa {\em et~al.},
  \ifthenelse{\boolean{articletitles}}{\emph{{High-statistics study of
  $\tau^-\to\pi^-\pi^0\nu_{\tau}$ decay}},
  }{}\href{https://doi.org/https://doi.org/10.1103/PhysRevD.78.072006}{Phys.\
  Rev.\  \textbf{D78} (2008) 072006}\relax
\mciteBstWouldAddEndPuncttrue
\mciteSetBstMidEndSepPunct{\mcitedefaultmidpunct}
{\mcitedefaultendpunct}{\mcitedefaultseppunct}\relax
\EndOfBibitem
\bibitem{paper:babar-isr}
BaBar collaboration, J.~P. Lees {\em et~al.},
  \ifthenelse{\boolean{articletitles}}{\emph{{Precise measurement of $e^+e^-\to
  \pi^+\pi^-(\gamma)$ with initial state radiations method at BaBar}},
  }{}\href{https://doi.org/https://10.1103/PhysRevD.86.032013}{Phys.\ Rev.\
  \textbf{D86} (2012) 032013}\relax
\mciteBstWouldAddEndPuncttrue
\mciteSetBstMidEndSepPunct{\mcitedefaultmidpunct}
{\mcitedefaultendpunct}{\mcitedefaultseppunct}\relax
\EndOfBibitem
\bibitem{pelaez19}
J.~R. Pelaez, A.~Rodas, and J.~Ruiz~de Elvira,
  \ifthenelse{\boolean{articletitles}}{\emph{{Global parameterization of
  $\pi\pi$ scattering up to 2 \gev}},
  }{}\href{https://doi.org/https://doi.org/10.1140/epjc/s10052-019-7509-6}{Eur.\
  Phys.\ J.\  \textbf{C79} (2019) 1008},
  \href{http://arxiv.org/abs/1907.13162}{{\normalfont\ttfamily
  arXiv:1907.13162}}\relax
\mciteBstWouldAddEndPuncttrue
\mciteSetBstMidEndSepPunct{\mcitedefaultmidpunct}
{\mcitedefaultendpunct}{\mcitedefaultseppunct}\relax
\EndOfBibitem
\end{mcitethebibliography}

 
\newpage
\centerline
{\large\bf LHCb collaboration}
\begin
{flushleft}
\small
R.~Aaij$^{32}$\lhcborcid{0000-0003-0533-1952},
A.S.W.~Abdelmotteleb$^{50}$\lhcborcid{0000-0001-7905-0542},
C.~Abellan~Beteta$^{44}$,
F.~Abudin{\'e}n$^{50}$\lhcborcid{0000-0002-6737-3528},
T.~Ackernley$^{54}$\lhcborcid{0000-0002-5951-3498},
B.~Adeva$^{40}$\lhcborcid{0000-0001-9756-3712},
M.~Adinolfi$^{48}$\lhcborcid{0000-0002-1326-1264},
P.~Adlarson$^{77}$\lhcborcid{0000-0001-6280-3851},
H.~Afsharnia$^{9}$,
C.~Agapopoulou$^{13}$\lhcborcid{0000-0002-2368-0147}
% ...

{\footnotesize \it

% ...
$^{9}$Universit{\'e} Clermont Auvergne, CNRS/IN2P3, LPC, Clermont-Ferrand, France\\
% ...
$^{13}$LPNHE, Sorbonne Universit{\'e}, Paris Diderot Sorbonne Paris Cit{\'e}, CNRS/IN2P3, Paris, France\\
% ...
$^{32}$Nikhef National Institute for Subatomic Physics, Amsterdam, Netherlands\\
% ...
$^{40}$Instituto Galego de F{\'\i}sica de Altas Enerx{\'\i}as (IGFAE), Universidade de Santiago de Compostela, Santiago de Compostela, Spain\\
% ...
$^{44}$Physik-Institut, Universit{\"a}t Z{\"u}rich, Z{\"u}rich, Switzerland\\
% ...
$^{48}$H.H. Wills Physics Laboratory, University of Bristol, Bristol, United Kingdom\\
% ...
$^{50}$Department of Physics, University of Warwick, Coventry, United Kingdom\\
% ...
$^{53}$School of Physics and Astronomy, University of Glasgow, Glasgow, United Kingdom\\
$^{54}$Oliver Lodge Laboratory, University of Liverpool, Liverpool, United Kingdom\\
% ...
$^{77}$Department of Physics and Astronomy, Uppsala University, Uppsala, Sweden, associated to $^{53}$\\
% ...
}
\end{flushleft}


\end{document}